\documentclass[lettersize,journal]{IEEEtran}
\usepackage{amsmath,amsfonts}
\usepackage{algorithmic}
\usepackage{algorithm}
\usepackage{array}
\usepackage{textcomp}
\usepackage{stfloats}
\usepackage{url}
\usepackage{verbatim}
\usepackage{graphicx}
\usepackage{cite}
\usepackage{cleveref}
\usepackage{amsthm}

\usepackage{bbm}
\usepackage{dsfont}
\usepackage{xcolor}

\newtheorem{theorem}{Theorem}
\newtheorem{lemma}{Lemma}
\newtheorem{proposition}{Proposition}

\newtheorem{definition}{Definition}
\newtheorem{remark}{Remark}

\usepackage{subcaption}
\usepackage{balance}

\newcommand*{\Scale}[2][4]{\scalebox{#1}{$#2$}}%

\begin{document}

\title{To Optimize Human-in-the-loop Learning in Repeated Routing Games}

\author{Hongbo Li,~\IEEEmembership{Member,~IEEE} and Lingjie Duan,~\IEEEmembership{Senior Member,~IEEE}

\thanks{This work is also supported in part by the Ministry of Education, Singapore, under its Academic Research Fund Tier 2 Grant with Award no. MOE-T2EP20121-0001; in part by SUTD Kickstarter Initiative (SKI) Grant with no. SKI 2021\_04\_07; and in part by the Joint SMU-SUTD Grant with no. 22-LKCSB-SMU-053.}

\thanks{Hongbo Li and Lingjie Duan are with the Pillar of Engineering Systems and Design, Singapore University of Technology and Design, Singapore 487372 (e-mail: hongbo\_li@mymail.sutd.edu.sg; lingjie\_duan@sutd.edu.sg).}
}

\markboth{IEEE Transactions on Mobile Computing}%
{Shell \MakeLowercase{\textit{et al.}}: A Sample Article Using IEEEtran.cls for IEEE Journals}

\IEEEpubid{0000--0000/00\$00.00~\copyright~2021 IEEE}

\maketitle

\begin{abstract}
Today navigation applications (e.g., Waze and Google Maps) enable human users to learn and share the latest traffic observations, yet such information sharing simply aids selfish users to predict and choose the shortest paths to jam each other. Prior routing game studies focus on myopic users in oversimplified one-shot scenarios to regulate selfish routing via information hiding or pricing mechanisms. 
For practical human-in-the-loop learning (HILL) in repeated routing games, we face non-myopic users of differential past observations and need new mechanisms (preferably non-monetary) to persuade users to adhere to the optimal path recommendations. 
We model the repeated routing game in a typical parallel transportation network, which generally contains one deterministic path and $N$ stochastic paths. 
We first prove that no matter under the information sharing mechanism in use or the latest routing literature's hiding mechanism, the resultant price of anarchy (PoA) for measuring the efficiency loss from social optimum can approach infinity, telling arbitrarily poor exploration-exploitation tradeoff over time. Then we propose a novel user-differential probabilistic recommendation (UPR) mechanism to differentiate and randomize path recommendations for users with differential learning histories. 
We prove that our UPR mechanism ensures interim individual rationality for all users and significantly reduces $\text{PoA}=\infty$ to close-to-optimal $\text{PoA}=1+\frac{1}{4N+3}$, which cannot be further reduced by any other non-monetary mechanism. In addition to theoretical analysis, we conduct extensive experiments using real-world datasets to generalize our routing graphs and validate the close-to-optimal performance of UPR mechanism.
\end{abstract}


\section{Introduction}\label{section1}
With ever-increasing smart devices owned by human users in transportation networks, mobile crowdsourcing applications (e.g., Waze and Google Maps) have emerged as vital platforms for users to learn and share their observed time-varying traffic information for guiding their daily or repeated routing (\!\!\cite{waze,olariu2021vehicular,kang2024metaverses}). 
These applications simply publicize all useful traffic information, which encourages selfish users to consistently choose the current shortest routes in each round to minimize their own travel costs (\!\!\cite{vasserman2015implementing,hasan2016multiagent}). As a consequence, users miss critical exploration of non-shortest paths that may turn better to route and reduce travel costs in the future. 

In systems characterized by dynamic and diverse information dynamics, Multi-Armed Bandit (MAB) problems emerge as a framework for investigating optimal exploration-exploitation trade-offs among stochastic arms or paths (e.g., \cite{slivkins2019introduction,li2020multi,gupta2021multi,bozorgchenani2021computation,wang2022truthful,li2023collaborative}). For instance, \cite{bozorgchenani2021computation} utilizes MAB techniques to forecast congestion levels in rapidly changing transportation environments. Similarly, \cite{wang2022truthful} employs MAB exploration-exploitation strategies to manage the recruitment of previously reliable users alongside new, unproven users in completing crowdsourcing tasks.
More recently, research efforts have extended traditional MAB paradigms to distributed settings involving multiple non-myopic agents aiming to minimize overall social costs (e.g., \cite{shi2021federated,yang2021cooperative,yang2022distributed,zhu2023distributed}). For instance, \cite{shi2021federated} explores the use of distributed agents to learn heterogeneous local models associated with each arm, contributing to the social planner's decision-making process. Additionally, \cite{yang2021cooperative} investigates how a fixed flow of agents collaborates in a distributed MAB scenario over repeated interactions, where each agent autonomously makes optimal decisions based on privately acquired arm information.
Subsequently, \cite{yang2022distributed} and \cite{zhu2023distributed} allow agents to learn and share the arm information with their neighbors for making locally optimal decisions. Overall, these new distributed MAB solutions require all agents to be fully cooperative to adhere to the optimal policy set by the social planner, without any possible selfish deviation to favor individuals' rewards.

\IEEEpubidadjcol

To facilitate human-in-the-loop learning (HILL) in repeated routing games, it is crucial to incentivize selfish users to follow the social planner's optimal recommendations, enabling both information learning on diverse paths and the long-term reduction of traffic congestion. 
Prior routing game studies mainly focus on an oversimplified one-shot routing scenario without any information variation over time. They develop information hiding (e.g., \cite{mazurczyk2016information,tavafoghi2017informational,wu2019information}) and side-payment or pricing mechanisms (e.g., \cite{ferguson2021effectiveness,yue2021incentive,li2022online}) to regulate myopic users' selfish routing choices. 
However, these mechanisms strongly assume that the social planner possesses all the traffic information to design incentives. Moreover, they only consider persuading myopic users to follow, whereas myopic users do not have past traffic observations as side information to reverse-engineer and deviate. 
In contrast, our HILL problem deals with a dynamic system with time-varying traffic information, necessitating the social planner to not only reduce immediate congestion caused by users but also incentivize their ongoing information learning on diverse paths. 

As compared to pricing or side-payment, informational mechanisms (e.g., information hiding) are non-monetary and easy to implement (\!\!\cite{fudenberg1991game,borgers2015introduction,li2019recommending}). In this paper, we aim to design the best informational mechanism to regulate HILL in repeated routing games to persuade non-myopic users with differential side information in routing choices. As such, we need to address the following two technical challenges. 
\begin{itemize}
    \item The first question is how to \emph{evaluate existing routing policies' performances for non-myopic users as compared to the social optimum in the dynamic system with time-varying traffic information}. Prior routing game literature (e.g. \cite{tavafoghi2017informational,wu2019information}) focuses on myopic users in oversimplified one-shot scenarios, assuming that the social planner knows all traffic conditions to decide routing policies. 
    In practice, the social planner lacks access to live traffic conditions but relies on non-myopic users to explore different paths over time. Although non-myopic, users remain selfish, and their interests are not aligned with the social planner's goal of balancing long-term exploitation and exploration to minimize the overall social cost. 
    Moreover, the theoretical analysis of the existing routing policies is lacking, particularly in understanding their performance gaps from the social optimum. 
    \item The second challenge is how to \emph{persuade non-myopic users to follow the optimal path recommendations despite their selfishness and differential learning histories to reverse-engineer}. In repeated routing games, each user only cares about its own travel cost. It may learn from its own traffic observations in the past to infer the current system's status and deviate from the optimal recommendation. Existing studies of incentivized exploration (\!\!\cite{kremer2014implementing,mansour2022bayesian,hu2022incentivizing,li2023congestion}) and Bayesian persuasion (\!\!\cite{kamenica2011bayesian,das2017reducing,farhadi2022dynamic,bernasconi2023optimal}) largely assume that all users are myopic to participate and do not have private side information to play strategically against the system's recommendations. 
\end{itemize}

It is worth noting that there is a recent study related to this paper (\!\!\cite{li2024human}), which introduces an informational mechanism aimed at regulating myopic users' distributed information learning in congestion games. 
However, for practical human-in-the-loop learning in repeated routing games, we face non-myopic users, who focus on their own long-term rewards rather than one-shot immediate rewards. In this case, the mechanism in \cite{li2024human} cannot work, as non-myopic users can reverse-engineer the system status to play against the mechanism.
Unlike this work, our paper addresses the two challenges above by proposing new mechanism designs and analysis.

We summarize our key novelty and main contributions as follows.
\begin{itemize}
    \item \emph{To regulate human-in-the-loop learning (HILL) for repeated routing games:} To the best of our knowledge, this work is the first to regulate non-myopic users with differential side-information in repeated routing games. We model the repeated routing game in a typical parallel transportation network, which generally contains one deterministic path and $N$ stochastic paths. Here each stochastic path has time-varying unknown traffic conditions to be learned by repeatedly arriving non-atomic users. Furthermore, users' past routing choices create differential side information for selfish routing, making it challenging to design any informational mechanism. 
    \item \emph{Theoretical comparison among information sharing, hiding mechanisms, and social optimum via PoA analysis:} We analyze the information sharing and hiding mechanisms as well as social optimum to derive their corresponding routing recommendations in closed form, respectively. Then we prove that no matter under today's information sharing mechanism (in Waze and Google Maps) or the latest information hiding mechanism in the routing game literature, the resultant price of anarchy (PoA) for measuring the efficiency loss from optimum about total users’ travel costs can be arbitrarily large.
    \item \emph{Best informational mechanism to remedy huge efficiency loss:} We propose a novel user-differential probabilistic recommendation (UPR) mechanism to randomly differentiate path recommendations for users with differential learning histories. We prove that our UPR mechanism ensures interim individual rationality for all users and significantly reduces $\text{PoA}=\infty$ to close-to-optimal $\text{PoA}=1+\frac{1}{4N+3}$, which cannot be further reduced by any other informational mechanism. In addition to theoretical analysis, we conduct extensive experiments using real-world datasets to generalize our routing graphs and validate the close-to-optimal average performance of our UPR mechanism.
\end{itemize}

The rest of the paper is organized as follows. In \Cref{section2}, we propose the repeated routing game model, the public human-in-the-loop learning (HILL) model under information sharing mechanism, and the differential HILL model under information hiding mechanism. Then in \Cref{section3}, we first formulate the optimization problems for the sharing and hiding mechanisms, as well as the socially optimal policy, and then analyze the solutions for these three policies. In \Cref{section4}, we analytically compare the two mechanisms with the social optimum via PoA analysis. Based on the analysis, we propose our UPR mechanism in \Cref{section5}. We verify the average performance of our UPR mechanism in \Cref{section6}. Finally, \Cref{section7} concludes the paper. To facilitate readability, we have summarized all the key notations in \Cref{notation_table}.

\section{System model}\label{section2}

\begin{figure}[t]
    \centering
    \begin{subfigure}{\linewidth}
        \centering
        \includegraphics[width=0.7\textwidth]{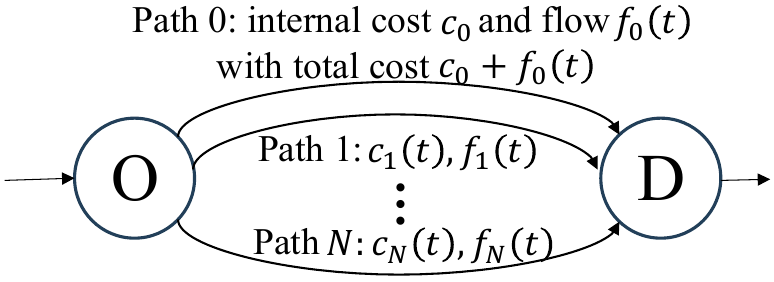}
        \caption{A typical parallel transportation network with one deterministic path and $N$ stochastic paths.}
        \label{fig:congestion_game}
    \end{subfigure}
    \begin{subfigure}{\linewidth}
        \centering\includegraphics[width=0.75\textwidth]{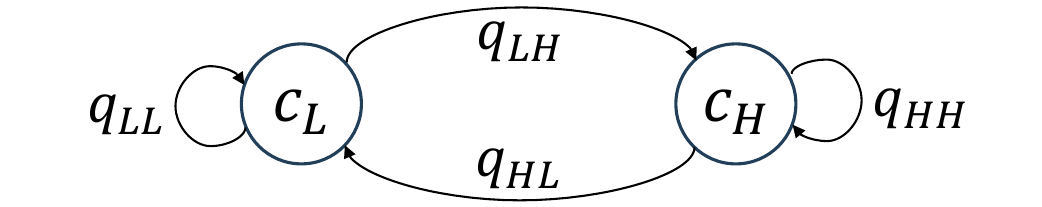}
        \caption{The Markov chain for modelling internal travel cost $c_i(t)$ dynamics of stochastic path $i\in\mathbb{N}=\{1,\cdots,N\}$ to alternate between low cost $c_L$ and high cost $c_H$, where $q_{LH}<q_{LL}$ and $q_{HL}<q_{HH}$.}
        \label{fig:mdp}
    \end{subfigure}
    \caption{At the beginning of each $t\in\{1,2,\cdots\}$, a unit flow of users arrive at origin O to select a path among all the $N+1$ paths of the network in Fig. \ref{fig:congestion_game}. Here path 0 has a fixed internal travel cost $c_0$. Yet any other path $i\in\mathbb{N}$ is stochastic and its internal cost $c_i(t)$ varies based on the Markov chain illustrated in Fig.~\ref{fig:mdp}. }
    \label{pricing_fig}
\end{figure}
In this section, as in the routing game literature (e.g., \cite{tavafoghi2017informational,wu2019information,li2023congestion}), we examine a standard parallel transportation network, which is simple but fundamental, to introduce the repeated routing game model.
Then, we proceed to present the public HILL model under the information sharing mechanism and the differential HILL model under the information hiding mechanism for the crowdsourcing platform, respectively. 

\subsection{Repeated Routing Game Model}
We consider an infinite discrete time horizon $t \in \{0,1,\cdots\}$. In Figure \ref{fig:congestion_game}, at the beginning of each $t$, there is a unit flow of users arriving at origin O. 
Then this unit flow of users individually choose paths from all the $N+1$ available paths to travel to their common destination D. We generally consider a deterministic path 0 with a fixed internal travel cost $c_0$ other than $N$ stochastic paths as in \cite{tavafoghi2017informational,wu2019information,li2023congestion}. 

Following existing routing game literature (e.g., \cite{das2017reducing,kang2024tiny,li2019recommending}), for any stochastic path $i\in\mathbb{N}=\{1,\cdots,N\}$, the internal travel cost $c_i(t)\in\{c_L,c_H\}$ is random and dynamically alternates between a high-cost state $c_H$ and a low-cost state $c_L$ over time, according to the commonly used Markov chain in Figure \ref{fig:mdp}.\footnote{Besides the references to support, we also validate the suitability of this Markov chain model in Section \ref{section6}'s experiments.} 
The internal cost may refer to factors like travel time or fuel consumption, which depend on the condition of each path (e.g., poor visibility, `black ice' segments \cite{li2023congestion,li2019recommending}).
Upon reaching destination D, users repeat the same process in the subsequent time slot $t+1$. The length of a time slot can represent a day, for instance, when people commute from home to the office each morning within a similar time range. Therefore, we generally assume that users are non-myopic, allowing them to retain their prior traffic observations to guide future routing decisions.

Given the unit flow of user arrivals at current $t$, we denote by $f_i(t)\in[0,1]$ the flow of users choosing any stochastic path~$i$. 
Following the existing literature (e.g., \cite{tavafoghi2017informational,das2017reducing,wu2019information}), in addition to the internal travel cost $c_i(t)$, each user on path $i$ will face another (external) congestion cost $f_i(t)$ caused by the other users on the same path. 
Therefore, the individual cost of each user traveling on stochastic path $i\in\mathbb{N}$ is $c_i(t)+f_i(t)$. Similarly, each of the other $f_0(t)$ flow of users on deterministic path 0 has individual cost $c_0+f_0(t)$, where
\begin{align*}
    f_0(t)=1-\sum_{i=1}^{N}f_i(t).
\end{align*}

To learn the time-varying internal cost $c_i(t)$ of stochastic path $i\in\mathbb{N}$ for future use, the platform (e.g., Waze) expects users to frequently travel on diverse paths and share their observations. 
Then it observes user flow $f_i(t)$ and observation $c_i(t)$ via GPS and travel time \cite{meigs2020optimal}.

\begin{table}[!t]
\renewcommand{\arraystretch}{1.3}
\caption{Key notations and their meanings in this paper}
\label{notation_table}
\centering
\begin{tabular}{|c|m{0.33\textwidth}|}
\hline
\textbf{Notation} & \textbf{Meaning}\\
\hline
\hline
$N$ & The number of stochastic paths.\\
\hline
$c_i(t)$ & The internal travel cost of stochastic path $i$.\\
\hline
$c_0(t)$ & The fixed internal travel cost of deterministic path 0.\\
\hline
$c_L, c_H$ & The high- and low-cost states on stochastic paths.\\
\hline
$q_{LH},q_{HH}$ & The transition probabilities from low- and high-cost states to high-cost state in the Markov chain.\\
\hline
$f_i(t)$ & The flow of users traveling any path $i$ at time $t$.\\
\hline
$x_i(t),x_i'(t)$ & The prior and posterior public hazard beliefs on stochastic path $i$ under the sharing mechanism.\\
\hline
$\mathbf{x}(t)$ & The set of public hazard beliefs of all stochastic paths.\\
\hline
$d_i(t)$ & The information age of path $i$ under the hiding mechanism.\\
\hline
$y_{i,d_i(t)}(t)$ & The private hazard belief with information age $d_i(t)$ on stochastic path $i$ under the hiding mechanism.\\
\hline
$\mathbf{y}(t)$ & The set of private hazard beliefs of all stochastic paths.\\
\hline
$c(\cdot)$ & The immediate social cost at time $t$.\\
\hline
$C(\cdot)$ & The long-term cost function for all users.\\
\hline
$\rho$ & The discount factor. \\
\hline
$\alpha$ & The expected arrival rate of users.\\
\hline
$\pi(t)$ & The private recommendation for each user under our UPR mechanism at time $t$.\\
\hline
\end{tabular}
\end{table}

\subsection{Public HILL under Information Sharing}\label{subsection2B}

The HILL model depends on the implemented mechanism, and we first consider the existing information-sharing mechanism used by Google Maps and Waze to share all useful traffic information with the public. 
After all users making routing decisions, a flow $f_i(t)$ of users travel on stochastic path $i$ to observe the actual state $c_i(t)$ there. 
We define a public hazard belief $x_i(t)\in[0,1]$ to be the expected probability for seeing the bad traffic or high-cost condition $c_i(t)=c_H$ at time $t$. Using the Bayesian inference, we obtain:
\begin{align}
    x_i(t)=\mathbf{Pr}(c_i(t)=c_H|c_i(t-1),x_i(t-1)),\label{def_x}
\end{align}
where $x_i(t-1)$ summarizes prior public hazard beliefs before $t-1$.
Then we introduce the information learning process of this public HILL model below.
\begin{itemize}
    \item At the beginning of time $t$, the platform broadcasts the public hazard belief $x_i(t)$ of any stochastic path $i$ in (\ref{def_x}) about traffic state $c_i(t)$.
    \item During time slot $t$, users traveling on stochastic path $i$ will observe the actual state $c_i(t)$ and report it back to the platform. Then the platform will further update this prior belief to a posterior belief $x_i'(t)$ below:
    \begin{itemize}
        \item If the $f_i(t)$ flow of users observe $c_i(t)=c_H$ on path $i$, then the posterior belief is updated to $x'_i(t)=1$.
        \item If these users observe $c_i(t)=c_L$ on path $i$, then the posterior belief is updated to $x'_i(t)=0$.
        \item If $f_i(t)=0$ without user to travel and learn this path~$i$, the posterior belief $x'_i(t)$ remains as $x_i(t)$. 
    \end{itemize}
    \item At the end of each time slot, the platform updates public hazard belief from $x_i'(t)$ to $x_i(t+1)$ below, based on the Markov chain in Fig. \ref{fig:mdp}: 
    \begin{equation}
        x_i(t+1)=x_i'(t)q_{HH}+\big(1-x_i'(t)\big)q_{LH}.\label{x(t+1)}
    \end{equation}
\end{itemize}
The above process will repeat as the new time slot $t+1$ begins.
Note that the Markov chain-based HILL process described above is equivalent to the existing restless multi-armed bandit (MAB) problem (e.g., \cite{tekin2012online,wang2020restless}). Our subsequent analysis and key results (e.g, \Cref{prop:PoA_m}, \Cref{prop:PoA_empty}, and our UPR mechanism) can also be generalized if $c_i(t)$ follows a memoryless stochastic process (e.g., Poisson process or exponential distribution), as the worst-case scenario remains unchanged.

\subsection{Differential HILL under Information Hiding}\label{subsection2C}
If the platform chooses to hide this public information $x_i(t)$ from users (e.g., by applying the latest information hiding mechanism \cite{tavafoghi2017informational,farhadi2022dynamic,li2019recommending}), users have to independently update private beliefs based on their own routing choices and observations of any path $i$ in the past. In this subsection, we introduce how users update their differential private beliefs without public information sharing by the platform. 
To aid users' routing decisions, we allow them to observe the traffic flow in the last time slot (i.e., $f_i(t-1)$ on any path~$i$) at the beginning of each~$t$, as widely assumed in existing transportation literature (e.g., \cite{castillo2008observability,salari2019optimization,he2024generative}).

At the initial time $t=0$, the hazard belief $x_i(0)$ of any stochastic path $i$ is given to all users. Without knowing public belief $x_i(t)$ but only the past flow $f_i(t-1)$ since $t=1$, users will experience different information ages since their last observations of this path. 
Denote by $d_i(t)$ the information age of path $i$ for an individual user at time $t\geq 1$, and let $y_{i,d_i(t)}(t)$ denote its corresponding private hazard belief. Then, we proceed to introduce the update of $d_i(t)$ and $y_{i,d_i(t)}(t)$ under differential HILL in the following two cases.
\begin{itemize}
    \item If this user just explored stochastic path $i$ at $t-1$, its information age of this path is set to the minimum $d_i(t)=1$. Therefore, its current private hazard belief is exactly the same as the public belief of the platform:
    \begin{align}
        y_{i,1}(t)=x_i(t).\label{y_i(1)}
    \end{align}
    \item If this user did not explore stochastic path $i$ at time $t-1$, its information age $d_i(t)$ should increase by $1$ from $d_i(t-1)$. As it is able to observe the last flow $f_i(t-1)$ at the beginning of $t$, it will rectify its private belief $y_{i,d_i(t-1)}(t-1)$. For example, if this user observes more users traveling on path $i$ with $f_i(t-1)\geq f_i(t-2)$ in the last time slot, it will believe that this path had a good traffic condition with weak hazard belief $x_i(t-1)=q_{LH}$ according to Fig. \ref{fig:mdp}. Otherwise, it infers $x_i(t-1)=q_{HH}$ with strong hazard belief there. By rectifying the last private belief to the exact public belief $x_i(t-1)$ from the prior flow distribution, its information age keeps at $d_i(t)=2$ and does not grow over time. Thus, its private belief for time $t$ is updated to:
    \begin{align}
        y_{i,2}(t)&=\mathbb{E}[x_i(t)|x_i(t-1)]\notag\\
        &=x_i(t-1)q_{HH}+(1-x_i(t-1))q_{LH},\label{y_i(2)}
    \end{align}
    due to (\ref{x(t+1)}) and $x_i'(t-1)=x_i(t-1)$ without new observation.
\end{itemize}

According to the above analysis, we summarize the dynamics of a user's information age $d_i(t)$ of path $i$ below:
\begin{align}
    d_i(t)=\begin{cases}
        1, &\text{if this user explored path $i$ at }t-1,\\
        2, &\text{otherwise,}
    \end{cases}\label{di(t)}
\end{align}
with the self-learning update of private hazard belief $y_{i,1}(t)$ in (\ref{y_i(1)}) and $y_{i,2}(t)$ in (\ref{y_i(2)}), respectively.

\section{Problem Formulations and Analysis for Sharing, Hiding, and Social Optimum}\label{section3}
In this section, we first formulate optimization problems for guiding non-myopic users' repeated routing under sharing and hiding mechanisms, as well as the social optimum. Then, we solve these optimization problems to determine the routing flow in closed form per time for each mechanism/policy. 

We first define non-myopic users' selfish routing decisions in the following.
\begin{definition}[Selfish routing decision]\label{def:selfish_policy}
Under the selfish routing decision, each non-myopic user aims to maximize its own long-term $\rho$-discounted expected reward, where $\rho\in(0,1)$ is the discount factor.
\end{definition}

Based on \Cref{def:selfish_policy}, each user will make selfish routing decisions under both sharing and hiding mechanisms.

\subsection{Formulation and Equilibrium for Information Sharing}\label{section3a}
In this subsection, we consider the information-sharing mechanism (used by Waze) to cater to users' selfish interests such that users will not deviate from its recommendations.

We first summarize the dynamics of public hazard belief $x_i(t)$ for each stochastic path $i$ at any time $t$ in a vector below:
\begin{align}
   \mathbf{x}(t)=\{x_1(t),\cdots,x_N(t)\},
\end{align}
each of which can be obtained by (\ref{x(t+1)}). 
For the unit user flow at time $t$, the platform first shares the latest public hazard belief set $\mathbf{x}(t)$ with them. Then each non-myopic user will selfishly choose the path that minimizes its own long-run cost. 

Let $Q^{(s)}\big(\mathbf{x}(t)|f_i^{(s)}(\mathbf{x}(t))\big)$ denote the minimal long-run expected cost of a single user since time $t$, given the $f_i^{(s)}(\mathbf{x}(t))\in[0,1]$ flow of non-myopic users choosing path $i$ under the sharing mechanism with symbol $(s)$ in the superscript. For ease of explanation, we take the special case of $N=1$ stochastic path alongside the deterministic path 0 in a typical two-path network example to solve and intuitively explain $f_1^{(s)}(x_1(t))$ below. For an arbitrary number $N\geq 2$, we can similarly compute 
$f_i^{(s)}(\mathbf{x}(t))$ by balancing the expected costs of all the $N+1$ paths as in (\ref{fm(t)}).
\begin{lemma}\label{lemma:fm(t)}
Under information sharing, given the unit flow of non-myopic user arrivals at time $t$, the recommended flow of users for choosing stochastic path 1 at equilibrium is: \footnote{To realize equilibrium flow $f_1^{(s)}(x_1(t))=\epsilon$ in the second case, each user will be fairly chosen with probability $\epsilon$ to explore path 1, and it will not deviate because $Q^{(s)}(x_1(t)|\epsilon)\leq Q^{(s)}(x_1(t)|0)$ for reduced long-run cost.}
\begin{align}
    f_1^{(s)}(x_1(t))=\begin{cases} 
        \epsilon\cdot \mathds{1}\{Q^{(s)}(x_1(t)|\epsilon)\leq Q^{(s)}(x_1(t)|0)\}, \\ \quad\quad\quad\quad\ \ \text{if } \mathbb{E}[c_1(t)|x_1(t)]\geq c_0+1, \\
        1, \ \quad \quad\quad\text{ if }\mathbb{E}[c_1(t)|x_1(t)]\leq c_0-1,\\
        \frac{1}{2}+\frac{c_0-\mathbb{E}[c_1(t)|x_1(t)]}{2}, \quad\ \quad \text{ otherwise},
    \end{cases}\label{fm(t)}
\end{align}
and $f_0^{(s)}(x_1(t))=1-f_1^{(s)}(x_1(t))$ to path 1, where infinitesimal $\epsilon>0$, $\mathds{1}\{(\cdot)\}=1$ if $(\cdot)$ is true and $\mathds{1}\{(\cdot)\}=0$ otherwise.
\end{lemma}

The proof of \Cref{lemma:fm(t)} is given in Appendix~A of the supplemental material.
From the first case of (\ref{fm(t)}), even if the stochastic path~1's internal travel cost $\mathbb{E}[c_1(t)|x_1(t)]$ is larger than path 0's internal cost $c_0$ plus the maximum congestion $1$ there, non-myopic users may still choose path 1 with minimum flow $\epsilon$ to learn possible $c_L$ state for future use given $Q^{(s)}(x_1(t)|\epsilon)\leq Q^{(s)}(x_1(t)|0)$.
If the internal cost of path 0 is larger than the internal cost of path 1 plus the maximum congestion, all user flow will choose path 1 in the first case of (\ref{fm(t)}). Otherwise, the two paths' conditions are comparable and the user flow will partition to both paths to balance congestion there. 

In the last two cases of (\ref{fm(t)}), as there is always a positive $f_1^{(s)}(x_1(t))>0$ flow of users traveling on path 1 to learn fresh information $c_1(t)$ there, non-myopic users will myopically make routing decisions to minimize their immediate individual costs in the current time slot. 

Then we further analyze the long-term social cost under the sharing mechanism (with specific (\ref{fm(t)}) for $N=1$ case). Define $C^{(s)}(\mathbf{x}(t))$ and $c^{(s)}(\mathbf{x}(t))$ to be the long-term and immediate social cost functions. At each time slot $t$, the immediate social cost is obtained as:
\begin{align}
    c^{(s)}(\mathbf{x}(t))=&f_0^{(s)}(\mathbf{x}(t))\Big(c_0+f_0^{(s)}(\mathbf{x}(t))\Big)\label{c(f(t))}\\&+\sum_{i=1}^Nf_i^{(s)}(\mathbf{x}(t))\Big(\mathbb{E}[c_i(t)|x_i(t)]+f_i^{(s)}(\mathbf{x}(t))\Big),\notag
\end{align}
which includes the current travel cost of $f_0^{(s)}(\mathbf{x}(t))$ flow of users on deterministic path 0 and $f_i^{(s)}(\mathbf{x}(t))$ flow of users on any stochastic path $i\in \mathbb{N}$. 

After obtaining immediate social cost (\ref{c(f(t))}), we further formulate the long-term cost function $C^{(s)}(\mathbf{x}(t))$ as a Markov decision process (MDP), where the update of hazard belief $x_i(t+1)$ depends on current users' routing choices and observations of $c_i(t)$ on path $i$'s. Consequently, two cases arise for updating $x_i(t+1)$ at each time slot. If $f_i^{(s)}(\mathbf{x}(t))=0$ without user traveling on stochastic path $i$, $x_i'(t)$ remains as $x_i(t)$ and will be updated to $x_i(t+1)$ according to (\ref{x(t+1)}). If $f_i^{(s)}(\mathbf{x}(t))>0$ with users traveling on path $i$, then $x_i(t+1)$ will be updated to $q_{HH}$ or $q_{LL}$, depending on whether users' observation there is $c_i(t)=c_H$ or $c_i(t)=c_L$ in Fig. \ref{fig:mdp}. 

Based on the analysis above, we formulate the long-term $\rho$-discounted cost function under information sharing below:
\begin{align}
    C^{(s)}\big(\mathbf{x}(t)\big)=c^{(s)}(\mathbf{x}(t))+\rho C^{(s)}\big(\mathbf{x}(t+1)\big),
    \label{Cm(x)}
\end{align}
where $c^{(s)}(\mathbf{x}(t))$ is given in (\ref{c(f(t))}) and $\rho\in(0,1)$ is the discount factor. Selfish users will not choose any stochastic path $i$ with higher long-run cost, and thus $x_i(t)$ of this path may not be updated on time to be useful for the next time slot.

\subsection{Formulation and Equilibrium for Information Hiding}\label{section3b}
In this subsection, we formulate the long-term cost function under the latest hiding mechanism in the routing game literature (e.g., \cite{tavafoghi2017informational,farhadi2022dynamic,li2019recommending}). Note that their mechanisms cannot be directly applied to our repeated routing as users are non-myopic and no longer hold the same private belief. Thus we need to revise it to fit our differential HILL model as described in Section \ref{subsection2C}. Note that the hiding mechanism still caters to users' selfish interests to hide all information including $\alpha(t)$ and lets users themselves choose routing paths that minimize their own long-run costs under \Cref{def:selfish_policy}.

Recall in (\ref{di(t)}) that users under the hiding mechanism are different in their private information ages $d_i(t)=1$ or $2$ from their own last observations, resulting in differential private belief $y_{i,1}(t)$ in (\ref{y_i(1)}) or $y_{i,2}(t)$ in (\ref{y_i(2)}) about path $i$, respectively. We summarize the dynamics of users' private beliefs $y_{i,1}(t)$ and $y_{i,2}(t)$ among $N$ stochastic paths at time $t$ as:
\begin{align}
 \mathbf{y}(t)=\big\{&y_{1,1}(t),\cdots,y_{N,1}(t),\\ &y_{1,2}(t),\cdots,y_{N,2}(t)\big\}.\notag
\end{align}

Let $f_i^{\emptyset}(\mathbf{y}(t))\in[0,1]$ denote the flow of non-myopic users choosing path $i$ at time $t$ under the hiding mechanism with subscript $\emptyset$, which needs to specify users' path decisions under different information age $d_i(t)$'s. Denote by $f_{i,d_i(t)}^{\emptyset}(y_{i,d_i(t)}(t))$ the flow of users choosing path $i$ with the same information age $d_i(t)$, which satisfies
\begin{align}
    f_i^{\emptyset}(\mathbf{y}(t))=\sum_{j=1}^2 f_{i,j}^{\emptyset}(y_{i,j}(t)).\label{f_empty(t)}
\end{align}

Given $f_i^{\emptyset}(\mathbf{y}(t-1))$ flow of users exploring path $i$ in the last time, the current flow of users with $d_i(t)=1$ is thus $f_i^{\emptyset}(\mathbf{y}(t-1))$. 
As they have the latest information on this path with $d_i(t)=1$, they will infer other users' private belief $y_{i,2}(t)$ with $d_i(t)=2$ according to (\ref{y_i(2)}).
Thus, they can estimate the exact flow $f_{i,2}^{\emptyset}(y_{i,2}(t))$ of these users on path $i$, and then decide $f^{\emptyset}_{i,1}(y_{i,1}(t))$ to minimize their own long-run costs. 

While for the other $1-f_i^{\emptyset}(\mathbf{y}(t-1))$ flow of users with $d_i(t)=2$, they only hold delayed or aged belief $y_{i,2}(t)$ to decide $f^{\emptyset}_{i,2}(y_{i,2}(t))$ to minimize their expected long-run costs. 

To clearly tell non-myopic users' decision-making under the differential HILL model, we derive in the next lemma the closed-form expressions for how users decide $f_{1,1}^{\emptyset}(y_{1,1}(t))$ and $f_{1,2}^{\emptyset}(y_{1,2}(t))$ in the basic two-path network at the equilibrium as an example.

\begin{lemma}\label{lemma:f_empty}
Under the information-hiding mechanism, for the $f_1^{\emptyset}(\mathbf{y}(t-1))$ flow of users with minimum information age $d_1(t)=1$ since their last observations, at time $t$ they decide their routing flow $f^{\emptyset}_{1,1}(y_{1,1}(t))$ on stochastic path 1 under the information-hiding mechanism as:
\begin{align}
    f^{\emptyset}_{1,1}(y_{1,1}(t))=\min\Big\{&\max\{f^{(s)}_1(y_{1,1}(t))-f^{\emptyset}_{1,2}(y_{1,2}(t)),0\},\notag\\ &f_1^{\emptyset}(\mathbf{y}(t-1))\label{f_empty1(t)}\Big\},
\end{align}
where $f_1^{(s)}(y_{1,1}(t))$ is given by (\ref{fm(t)}). For the other $1-f_1^{\emptyset}(\mathbf{y}(t-1))$ flow of users with age $d_1(t)=2$, their routing flow is
\begin{align}
    f^{\emptyset}_{1,2}(y_{1,2}(t))=(1-f_1^{\emptyset}(\mathbf{y}(t-1)))f_1^{(s)}(y_{1,2}(t)),\label{f_empty2(t)}
\end{align}
where $f_1^{(s)}(y_{1,2}(t))$ is given in (\ref{fm(t)}). Then the total flow of users on path 1 is $f^{\emptyset}_{1,1}(y_{1,1}(t))+f^{\emptyset}_{1,2}(y_{1,2}(t))$. 
\end{lemma}

The proof of \Cref{lemma:f_empty} is given in Appendix~B of the supplemental material.
As compared to $f_1^{(s)}(x_1(t))$ under information sharing in (\ref{fm(t)}), users with $d_i(t)=2$ under information hiding have to use their private belief $y_{1,2}(t)$ to decide $f_{1,2}^{\emptyset}(y_{1,2}(t))$ in (\ref{f_empty2(t)}). Then users with $d_i(t)=1$ decide $f_{1,2}^{\emptyset}(y_{1,2}(t))$ in (\ref{f_empty1(t)}) to make the total flow on path $1$ approach to $f_1^{(s)}(y_{1,1}(t))$ under information sharing as much as possible.

Next, we similarly formulate the long-term $\rho$-discounted cost function under the hiding mechanism below:
\begin{align}\label{Cempty(x)}
    C^{\emptyset}(\mathbf{y}(t))=c^{\emptyset}(\mathbf{y}(t))+\rho C^\emptyset(\mathbf{y}(t+1)),
\end{align}
where the immediate social cost $c^{\emptyset}(\mathbf{y}(t))$ is similarly defined as (\ref{c(f(t))}) and the update of $\mathbf{y}(t+1)$ is given in (\ref{y_i(1)}) or (\ref{y_i(2)}), depending on the system belief sets $\mathbf{x}(t-1)$ and $\mathbf{x}(t)$.

\subsection{Formulation and Solution for Social Optimum}\label{section3c}

Recall that the sharing and hiding mechanisms that cater to users' selfish interests for persuading them to follow path recommendations in Lemmas \ref{lemma:fm(t)} and \ref{lemma:f_empty}. Differently, the socially optimal policy aims to centrally solve the optimal routing flow $f^*_i(t)\in[0,\alpha(t)]$ to minimize the long-run expected social cost for all users, by assuming users to be cooperative all the time.
Let $C^*(\mathbf{x}(t))$ denote the long-term $\rho$-discounted cost function under the socially optimal policy, given as:
\begin{align}
    C^*(\mathbf{x}(t))=\min_{f^*_i(t)\in[0,1]} c^*(\mathbf{x}(t))+\rho C^*(\mathbf{x}(t+1)),\label{C*(x)}
\end{align}
where $c^*(\mathbf{x}(t))$ is similarly defined as (\ref{c(f(t))}). If the platform asks all the unit flow of arriving users to choose path 0, there would be no observation on stochastic path $i$. As a result, the posterior belief $x_i'(t)$ remains as $x_i(t)$. However, if the platform instructs some users to travel on path $i$, $x_i'(t)$ can be timely updated to $0$ or $1$ and will be further updated to $x_i(t+1)$ by (\ref{x(t+1)}). 

In (\ref{C*(x)}), even if the expected travel cost $\mathbb{E}[c_i(t)|x_i(t)]$ on path~$i$ is much higher than others, the socially optimal policy may still recommend a small flow $\epsilon> 0$ of users to learn $x_i(t)$ on path~$i$. As $\epsilon\rightarrow 0$, this infinitesimal flow of users' immediate total cost $\epsilon(\mathbb{E}[c_i(t)|x_i(t)]+\epsilon)$ is negligible. At the same time, their learned information on path $i$ can help reduce future costs for all users. Therefore, the socially optimal policy always decides $f_i^*(t)\geq \epsilon$ for any stochastic path $i$. 

Next, we derive the optimal recommended flow $f_1^*(x_1(t))$ in the typical two-path network example.
\begin{lemma}\label{Lemma:f*(t)}
Under the socially optimal policy, the recommended flow on stochastic path 1 is
\begin{align}
    f_1^*(x_1(t))=\begin{cases}
        \epsilon,\quad\quad \text{ if }\mathbb{E}[c_1(t)|x_1(t)]\geq c_0+2,\\
        1, \quad\quad \text{ if }\mathbb{E}[c_1(t)|x_1(t)]\leq c_0-2,\\
        \frac{1}{2}+\frac{c_0-\mathbb{E}[c_1(t)|x_1(t)]}{4}, \ \ \ \text{otherwise,}
    \end{cases}\label{f*(t)}
\end{align}
where $\epsilon>0$ is infinitesimal.
\end{lemma}

The proof of \Cref{Lemma:f*(t)} is given in Appendix~C of the supplemental material.
By comparing (\ref{f*(t)}) to (\ref{fm(t)}) under the sharing mechanism, we observe that the socially optimal policy always explores stochastic path 1 with $f_1^*(x_1(t))\geq \epsilon$, even if $\mathbb{E}[c_1(t)|x_1(t)]$ is much larger than $c_0$. If $\mathbb{E}[c_1(t)|x_1(t)]<c_0$, $f_1^*(x_1(t))=\frac{1}{2}+\frac{c_0-\mathbb{E}[c_1(t)|x_1(t)]}{4}$ in (\ref{f*(t)}) is smaller than $f_1^{(s)}(x_1(t))=\frac{1}{2}+\frac{c_0-\mathbb{E}[c_1(t)|x_1(t)]}{2}$ in (\ref{fm(t)}) to reduce the overall congestion on this stochastic path.

By comparing (\ref{f*(t)}) to (\ref{f_empty1(t)})-(\ref{f_empty2(t)}) under the hiding mechanism, the socially optimal policy adaptively uses actual $x_1(t)$ to guide users' routing in different cases.
However, the hiding mechanism uses $y_{1,1}(t)$ and $y_{1,2}(t)$ for partially blind routing decisions, which can be very different from the optimum. Based on (\ref{fm(t)}), (\ref{f_empty1(t)})-(\ref{f_empty2(t)}) and (\ref{f*(t)}) of the three mechanisms/policies, we are ready to compare their system performances in Section \ref{section4} to inspire our mechanism design.

\section{Analytical comparison between sharing, hiding, and social optimum via PoA}\label{section4}
In this section, we first conduct a comparative PoA analysis between the sharing mechanism and the socially optimal policy to prove the huge performance loss of the sharing mechanism. Subsequently, we also prove the huge PoA of the hiding mechanism as compared to the social optimum. Then we are well motivated to propose our new mechanism in Section~\ref{section5}. 
Actually, this section's analytical results provide important guidance to our later mechanism design, which is carefully designed to play between information-sharing and information-hiding mechanism. 

Before analyzing the two mechanisms' performance loss, we first obtain the bounds of public hazard belief $x_i(t)$ and private belief $y_{i,d_i(t)}(t)$ for future use.
\begin{lemma}\label{lemma:x(t+1)}
Given any initial belief $x_i(0)\in[0,1]$ for stochastic path $i$, both the public belief $x_i(t)$ and private belief $y_{i,d_i(t)}(t)$ satisfy $x_i(t)\in[q_{LH},q_{HH}]$ and $y_{i,d_i(t)}(t)\in[q_{LH},q_{HH}]$ for any time $t\geq 1$.
\end{lemma}

The proof of \Cref{lemma:x(t+1)} is given in Appendix~D of the supplemental material. According to (\ref{x(t+1)}), the lower bound $q_{LH}$ and upper bound $q_{HH}$ are derived when $x'(t)=0$ and maximum $x'(t)=1$, respectively.
If the current users observe $c_i(t)=c_L$, leading to $x'(t)=0$, the hazard belief $x_i(t+1)$ will be updated to its minimum value $q_{LH}$. Conversely, if $x_i'(t)=1$ with $c_i(t)=c_H$, $x_i(t+1)$ will be updated to its maximum value $q_{HH}$. Consequently, the updated hazard belief $x_i(t+1)$ will always remain bounded within $[q_{LH},q_{HH}]$. 
Similarly, based on the updates of the private belief $y_{i,d_i(t)}(t)$ in (\ref{y_i(1)}) or (\ref{y_i(2)}), it is also constrained by $q_{LH}$ and $q_{HH}$.

Based on Lemma \ref{lemma:x(t+1)}, if the minimal expected travel cost $\mathbb{E}[c_i(t)|x_i(0)=q_{LH}]$ under the good traffic condition of stochastic path $i$ is larger than $c_0$, selfish users will never explore stochastic path $i$ at any time. On the other hand, if the expected maximal expected travel cost $\mathbb{E}[c_i(t)|x_i(t)=q_{HH}]$ is smaller than $c_0$, selfish users will always choose path $i$ to exploit the low travel cost $c_0$ there. Therefore, 
To avoid the trivial cases of always exploration or exploitation, we focus on $\mathbb{E}[c_i(t)|q_{LH}]< c_0$, $\mathbb{E}[c_i(t)|q_{HH}]> c_0$, and $\mathbb{E}[c_i(0)|x_i(0)]< c_0+1$ below.

\subsection{PoA Analysis of Information Sharing Mechanism}
We compare $f_i^{(s)}(\mathbf{x}(t))$ under the sharing mechanism in Section \ref{section3a} to the optimal recommended flow $f_i^*(\mathbf{x}(t))$ under the social optimum in Section \ref{section3c} to examine the sharing mechanism's efficiency loss. For example, by comparing $f_1^{(s)}(\mathbf{x}(t))$ in (\ref{fm(t)}) and $f_1^*(\mathbf{x}(t))$ in (\ref{f*(t)}) for the two-path network example, we find that the sharing mechanism misses both exploitation and exploration on stochastic path~1, as compared to the social optimum. Below, we examine their performance gap for an arbitrary number $N$ of stochastic paths.

We use the standard definition of the price of anarchy (PoA) in game theory \cite{koutsoupias1999worst}, for measuring the maximum ratio between the long-term social cost under the sharing mechanism in (\ref{Cm(x)}) and the minimal social cost in (\ref{C*(x)}):
\begin{align}
    &\text{PoA}^{(s)}=\max_{\substack{c_0,c_H,c_L, q_{LL},\\q_{HH},\mathbf{x}(t),\rho}}\frac{C^{(s)}(\mathbf{x}(t))}{C^{*}(\mathbf{x}(t))},\label{def_PoA}
\end{align}
which is always larger than $1$ by considering all system parameters. 

Through involved worst-case analysis, we prove the lower bound of $\text{PoA}^{(s)}$ in (\ref{def_PoA}) caused by the sharing mechanism's lack of exploration in the following proposition.
\begin{proposition}\label{prop:PoA_m}
The PoA defined in (\ref{def_PoA}) under the information-sharing mechanism in (\ref{Cm(x)}) satisfies
\begin{align}\label{PoA_m}
    \text{PoA}^{(s)}\geq \frac{1-\rho+\rho\prod_{i=1}^N x_i(0)}{1-\rho+(\rho-\rho^{\frac{\lceil \log_{q_{HH}}\delta \rceil}{N}})\prod_{i=1}^N x_i(0)},
\end{align}
where $\delta$ is an infinitesimal.
This lower bound is tight as $q_{LL}\rightarrow 1, q_{HH}\rightarrow 1, c_L=0, c_H\gg c_0, \frac{1-q_{LL}}{1-q_{HH}}\rightarrow 0, \mathbb{E}[c_i(0)|x_i(0)]\rightarrow c_0+1$. 
\end{proposition}

The proof of \Cref{prop:PoA_m} is given in Appendix~E of the supplemental material. In \Cref{prop:PoA_m}'s proof, we consider the worst case of minimum exploration under the sharing mechanism. Given $\mathbb{E}[c_i(0)|x_i(0)]$ up to $c_0+1$ on path~$i$, there will be an $\epsilon$ flow of users traveling on stochastic path $i$ at time $t=0$. Once they find $c_i(t)=c_H$ there at any $t\geq 1$, no user will explore this path again, as $q_{HH}\rightarrow 1$.
However, socially optimal policy (\ref{f*(t)}) always recommends $\epsilon$ flow of users to explore any path $i$ to learn possible $c_i(t)=c_L$ there. As $\frac{1-q_{LL}}{1-q_{HH}}\rightarrow 0$, it takes them at most $\frac{\lceil \log_{q_{HH}}\delta \rceil}{N}$ time slots to find at least one path with $c_L=0$. After that, all users' long-term costs will be significantly reduced due to $q_{LL}\rightarrow 1$. Then we compare the two long-term costs to derive the PoA lower bound in (\ref{PoA_m}). 

\begin{remark}
In (\ref{PoA_m}), once $\mathcal{O}(\frac{1}{1-\rho})\gg \mathcal{O}(\frac{1}{1-q_{HH}})$ to make $\rho^{\frac{k}{N}}\rightarrow 1$, then $\text{PoA}^{(s)}=\infty$ to tell infinitely large efficiency loss under the sharing mechanism.
\end{remark}

In (\ref{PoA_m}), the PoA lower bound decreases with path number $N$, as more stochastic paths help risk pooling and make it easier for users to find $c_i(t)=c_L$ to approach the optimum. As $N\rightarrow \infty$ with infinite stochastic paths, $\text{PoA}^{(s)}$ in (\ref{PoA_m}) approaches the optimum $1$.

\subsection{PoA Analysis of Information Hiding Mechanism}
In this subsection, we further analyze the PoA lower bound caused by the hiding mechanism.
Let $\text{PoA}^{\emptyset}$ denote the PoA caused by the hiding mechanism, which is similarly defined as (\ref{def_PoA}) to compare the social cost $C^\emptyset(\mathbf{y}(t))$ in (\ref{Cempty(x)}) with $C^*(\mathbf{x}(t))$ in (\ref{C*(x)}). In the next proposition, we successfully provide the worst-case analysis to prove $\text{PoA}^{\emptyset}$'s lower bound.

\begin{proposition}\label{prop:PoA_empty}
As compared to the minimal social cost in (\ref{C*(x)}), the information-hiding mechanism in (\ref{Cempty(x)}) results in
\begin{align}\label{PoA_empty}
    \text{PoA}^{\emptyset}\geq \frac{1-\rho^2+\rho^2\prod_{i=1}^N x_i(0)}{1-\rho +(\rho-\rho^{\frac{\lceil \log_{q_{HH}}\delta \rceil}{N}})\prod_{i=1}^Nx_i(0)},
\end{align}
where $\delta$ is an infinitesimal. 
This lower bound is tight as $q_{LL}\rightarrow 1,q_{HH}\rightarrow 1, c_L=0, c_H\gg c_0, \frac{1-q_{LL}}{1-q_{HH}}\rightarrow 0$, $\mathbb{E}[c_i(0)|x_i(0)]\rightarrow c_0+1$. 
\end{proposition}

The proof of \Cref{prop:PoA_empty} is given in Appendix~F of the supplemental material. This PoA bound (\ref{PoA_empty}) can be similarly proved by considering the minimum exploration case as in \Cref{prop:PoA_m}. However, different from the sharing mechanism that always knows the latest system information, users with information delay $d_i(t)=2$ under the hiding mechanism have a time slot's latency to decide exploration or exploitation, which makes $\text{PoA}^{\emptyset}$ in (\ref{PoA_empty}) different from $\text{PoA}^{(s)}$ in (\ref{PoA_m})

\begin{remark}
In (\ref{PoA_empty}), once $\mathcal{O}(\frac{1}{1-\rho})\gg \mathcal{O}(\frac{1}{1-q_{HH}})$ to make $\rho^{\frac{k}{N}}\rightarrow 1$, then $\text{PoA}^{\emptyset}=\infty$, telling infinitely large efficiency loss under the information-hiding mechanism.
\end{remark}

In (\ref{PoA_empty}), the lower bound  of $\text{PoA}^{\emptyset}$ also decreases with path number $N$. As $N\rightarrow \infty$ with infinite paths, $\text{PoA}^{\emptyset}$ approaches $1+\rho$, due to users' late exploitations of stochastic paths to increase the total cost under information delay $d_i(t)=2$.

Given neither the sharing mechanism nor the hiding mechanism performs well for non-myopic users with finite $N$ in practice, it is critical to design a new mechanism other than simply sharing or hiding information.

\section{Design and Analysis of Our User-differential Informational Mechanism}\label{section5}
To motivate non-myopic users to follow the socially optimal policy in (\ref{f*(t)}) as much as possible, we aim to design the best mechanism to obtain the minimum possible PoA in this section. As pricing or side-payment mechanisms directly charge users, they are difficult to implement in the existing traffic navigation platforms (\!\!\cite{ferguson2021effectiveness,yue2021incentive,li2024incentivizing}). Instead, we aim to design a non-monetary informational mechanism that indirectly restricts information channels to regulate users.
Following the mechanism design literature (e.g., \cite{fudenberg1991game} and \cite{borgers2015introduction}), we provide its general definition below.
\begin{definition}[Informational mechanism]\label{def:informational_mechanism}
An informational mechanism defines a Bayesian game, where users send costless information and the social planner decides the disclosing level of observable variables based on received information. 
\end{definition}

By Definition \ref{def:informational_mechanism}, we notice that the information hiding mechanism in Section \ref{section3b} belongs to informational mechanisms, by simply hiding all traffic information from selfish users. Yet its achieved PoA is arbitrarily large in Proposition~\ref{prop:PoA_empty}. 
Additionally, we analyze existing deterministic recommendation mechanisms (e.g., \cite{tavafoghi2017informational, wu2019information, li2019recommending}), which offer state-dependent deterministic recommendations to users while concealing other information, and demonstrate that these mechanisms still result in an infinite PoA, motivating our UPR mechanism design.
\begin{lemma}\label{lemma:PoA_determinstic_mechanism}
The existing deterministic-recommendation mechanisms make $\text{PoA}=\infty$ in this system.
\end{lemma}
The proof of \Cref{lemma:PoA_determinstic_mechanism} is given in Appendix~G of the supplemental material. 
Intuitively, when a user receives recommendation $\pi(t)=i$, it still infers the expected travel cost for each path based on its private information $y_{i,d_i(t)}(t)$.
As a result, the recommendation $\pi(t)$ does not influence the user's selfish path choice, ultimately leading to an infinite PoA.

We next examine the efficiency limit of any informational mechanism for our HILL system below.
\begin{proposition}\label{prop:informational_mechanism}
In our dynamic HILL system, under any informational mechanism in \Cref{def:informational_mechanism}, the achieved PoA satisfies
\begin{align*}
    \text{PoA}\geq 1+\frac{1}{4N+3}.
\end{align*} 
\end{proposition}

The proof of \Cref{prop:informational_mechanism} is given in Appendix~H of the supplemental material. To obtain this minimal possible PoA, we first show that no informational mechanism in Definition \ref{def:informational_mechanism} can prevent users' maximum exploration of stochastic paths. Let $x_i(0)=q_{LH}, q_{LH}\rightarrow 0, q_{HH}<1, c_L=0$ for any stochastic path $i\in \mathbb{N}$. 
Given $\mathbb{E}[c_i(0)|x_i(0)]= c_0-\frac{1}{N}$, there will be $f_i^{(s)}(\mathbf{x}(0))=\frac{1}{N}$ flow of users traveling on each stochastic path~$i$ at initial $t=0$, according to (\ref{fm(t)}). 
Since initial hazard belief $x_i(0)=q_{LH}\rightarrow 0$, users will observe $c_i(0)=c_L$ with good traffic condition on path~$i$. 
Then each user is aware of the public belief $x_i(1)=q_{LH}$ on its chosen stochastic path~$i$ with minimum information age $d_i(1)=1$ for next time $t=1$. 
Consequently, they will behave the same as the sharing mechanism by continuing to travel to their chosen stochastic paths since $t= 1$, making any informational mechanism fail to regulate.

On the other hand, according to $f^*(\mathbf{x}(t))$ in (\ref{f*(t)}), the socially optimal policy ensures $\frac{1}{2(N+1)}$ flow of users to choose deterministic path 0 and avoid congestion cost on stochastic paths. In this case, the resulting minimal PoA is $1+\frac{1}{4N+3}$.

Building upon \Cref{prop:informational_mechanism}, we want to design the best informational mechanism to achieve $\text{PoA}=1+\frac{1}{4N+3}$. Based on the worst-case analysis in the proof above, when $c_i(t-1)=c_L$, no informational mechanism is able to prevent users with minimum information age $d_i(t)=1$ from exploiting stochastic path $i$ at the current time. When a number $1\leq n\leq N$ of stochastic paths reach their minimum public hazard belief (i.e., $x_i(t)=q_{LH}$ for path $i$ according to Lemma \ref{lemma:x(t+1)}), selfish users have the maximum tendency to exploit these paths and the resultant routing flow on path $i$ reaches its maximum: 
\begin{equation}
    \Bar{f}^{(s)}(n,t)=\min\left\{\frac{1}{n},\frac{c_0+1-\mathbb{E}[c_i(t)|x_i(t)=q_{LH}]}{n+1}\right\}.\label{bar_fm_i}
\end{equation}

Given the number $n$ of stochastic paths in good condition at time $t$, $\Bar{f}^{(s)}(n,t)$ is known to all users. 
Alternatively, when $c_i(t-1)=c_H$, no informational mechanism can persuade users with minimum age $d_i(t)=1$ to choose path $i$ at $t$. 
Hence, we can only focus on incentivizing the remaining users with non-trivial information age $d_i(t)=2$ to align with the platform's optimal path recommendations. 

On one hand, different from the hiding mechanism to blind all users, this user-differential approach inspires us to differentiate our path recommendations to users of differential $d_i(t)$. 
On the other hand, unlike the sharing mechanism, our mechanism design does not share with users the public belief set history $\{\mathbf{x}(1),\cdots,\mathbf{x}(t)\}$ but the previous flow distribution to improve efficiency. As users (with information age $d_i(t)=2$) can reverse-engineer from the system's path recommendations to infer the actual hazard belief, we propose probabilistic recommendations below. 

In the following mechanism, let $f^{(\text{UPR})}_i(t)\in[\epsilon,\Bar{f}^{(s)}(n,t)]$ represent the recommended flow of path $i$ at time $t$, which is always positive and keeps learning from all stochastic paths over time.

\begin{definition}[User-differential Probabilistic Recommendation (UPR) mechanism]\label{def:UPR}
At any given time $t\geq 1$, the platform operates in the following two steps:

\emph{Step 1: Selective information hiding.} First, the platform hides public belief history $\{\mathbf{x}(1),\cdots,\mathbf{x}(t)\}$ from all users but discloses previous routing flow (i.e., $f^{(\text{UPR})}_i(t-1)$ for any path~$i$). 

\emph{Step 2: Probabilistic Recommendation.}
Then the platform performs the following user-differential random signaling strategies to disclose selected information.
\begin{itemize}
    \item For the $f_i^{(\text{UPR})}(t-1)$ flow of users observing $c_i(t-1) = c_L$ on path $i$ with minimum age $d_i(t)=1$, the platform just recommends them to continue and make their own selfish routing decisions in (\ref{f_empty1(t)}).
    \item For the other users with $d_i(t)=2$ of any stochastic path $i\in I(t):=\{i\in \mathbb{N}|\Bar{f}^{(s)}(n,t)>f^{(\text{UPR})}_i(t-1)\}$, the platform independently recommends each of them to travel on path $\pi(t)=i$ with probability
\begin{align}\label{pr(pi=i)_multi}
    \!\!\!\!&\mathbf{Pr}(\pi(t)=i|x_i(t))=\\
     &\begin{cases}
        \epsilon,\quad\quad\quad\quad\quad\quad\ \ \ \text{ if high hazard belief }x_i(t)=q_{HH},\\
        \frac{{\Bar{f}^{(s)}(n,t)-f^{(\text{UPR})}_i(t-1)}}{1-f^{(\text{UPR})}_i(t-1)}, \text{if low hazard belief }x_i(t)=q_{LH}.\notag
    \end{cases}
\end{align}
\end{itemize}
    Besides, the platform independently recommends each of them to travel to deterministic path $\pi(t)=0$ with probability 
    \begin{align*}
        \mathbf{Pr}(\pi(t)=0|x_i(t))=1-\sum_{i\in I(t)}\mathbf{Pr}(\pi(t)=i|x_i(t)).
    \end{align*}
\end{definition}

Here, our UPR mechanism does not prevent users, who just observed $c_i(t-1)=c_L$ with minimum age $d_i(t)=1$, from pursuing selfish interests by exploiting stochastic path~$i$ at time $t$. However, for those users who just observed $c_j(t-1)=c_H$ on another path $j$ and other users who just traveled on deterministic path 0 with obvious age $d_i(t)=2$, they receive probabilistic path recommendations in (\ref{pr(pi=i)_multi}) to deviate from this strong hazard path and try other paths. 

Our mechanism proactively designs (\ref{pr(pi=i)_multi}) in the persuasive way that a user with $d_i(t)=2$ hearing the realized recommendation $\pi(t)=i$ will reverse-engineer path $i$ in a good condition $x_i(t)=q_{LH}$, as the corresponding recommendation probability $\frac{{\Bar{f}^{(s)}(n,t)-f^{(\text{UPR})}_i(t-1)}}{1-f^{(\text{UPR})}_i(t-1)}\gg \epsilon$ by comparing the two cases of (\ref{pr(pi=i)_multi}). On the other hand, if a user is recommended to path $\pi(t)=0$, it will similarly believe that any stochastic path $i$ has a bad traffic condition with $x_i(t)=q_{HH}$. As a result, it will follow this recommendation to choose path 0.

In summary, our UPR mechanism enables users to exploit $c_i(t)=c_L$ on stochastic path $i$, when the public hazard belief $x_i(t)=q_{LH}$ on this path is low. On the other hand, when stochastic path $i$ has a high hazard belief $x_i(t)=q_{HH}$, our UPR mechanism still ensures an $\epsilon$ flow of users exploring this path to learn for future. Thus, our UPR mechanism successfully avoids the worst-case scenarios with minimum exploration under both the sharing and hiding mechanisms.
Note that the main idea of our UPR mechanism can also be generated to regulate selfish players in other restless MAB games (e.g., \cite{tekin2012online,wang2020restless}). In these problems, the social planner provides differential recommendations to users based on their private information about each arm (path), which influences their posterior beliefs and consequently regulates their selfish arm choices.

Following \cite{cramton1987dissolving} and \cite{kosenok2008individually}, we define users' Interim Individual Rationality (IIR) to ensure their long-term participation in our UPR mechanism.
\begin{definition}[Interim Individual Rationality]\label{def:IIR}
A mechanism ensures Interim Individual Rationality for users with private information if, prior to the realization of the cost incurred by following the mechanism, each user anticipates achieving a cost level that is at least as low as their reservation cost.
\end{definition}

Based on above analysis and \Cref{def:IIR}, we prove our UPR mechanism ensures IIR for all users and achieves close-to-optimal performance in the following theorem.
\begin{theorem}\label{thm:UPR_PoA}
Our UPR mechanism in Definition \ref{def:UPR} ensures interim individual rationality for all non-myopic users to follow the system's random path recommendations. It greatly reduces PoA from arbitrarily large in (\ref{PoA_m}) and (\ref{PoA_empty}) to the minimum possible ratio: 
\begin{align}
    \text{PoA}^{(\text{UPR})}=1+\frac{1}{4N+3},\label{PoA_UPR}
\end{align}
which is always no larger than $\frac{8}{7}$ for any $N\geq 1$.
\end{theorem}

The proof of \Cref{thm:UPR_PoA} is given in Appendix~I of the supplemental material.
Recall \Cref{prop:informational_mechanism} tells that no informational mechanism can achieve PoA less than $1+\frac{1}{4N+3}$, it demonstrates that our UPR mechanism is the best to reduce PoA. The worst case under our UPR mechanism is the same as \Cref{prop:informational_mechanism}, which occurs at users' maximum exploration when all stochastic paths are in good traffic conditions with $x_i(t)=q_{LH}$.

In addition to the worst-case PoA analysis presented in Theorem \ref{thm:UPR_PoA}, we further conduct experiments using real datasets to examine the average performance of our UPR mechanism.

\begin{figure}[t]
    \captionsetup{font={footnotesize}}
    \centering
    \includegraphics[width=0.45\textwidth]{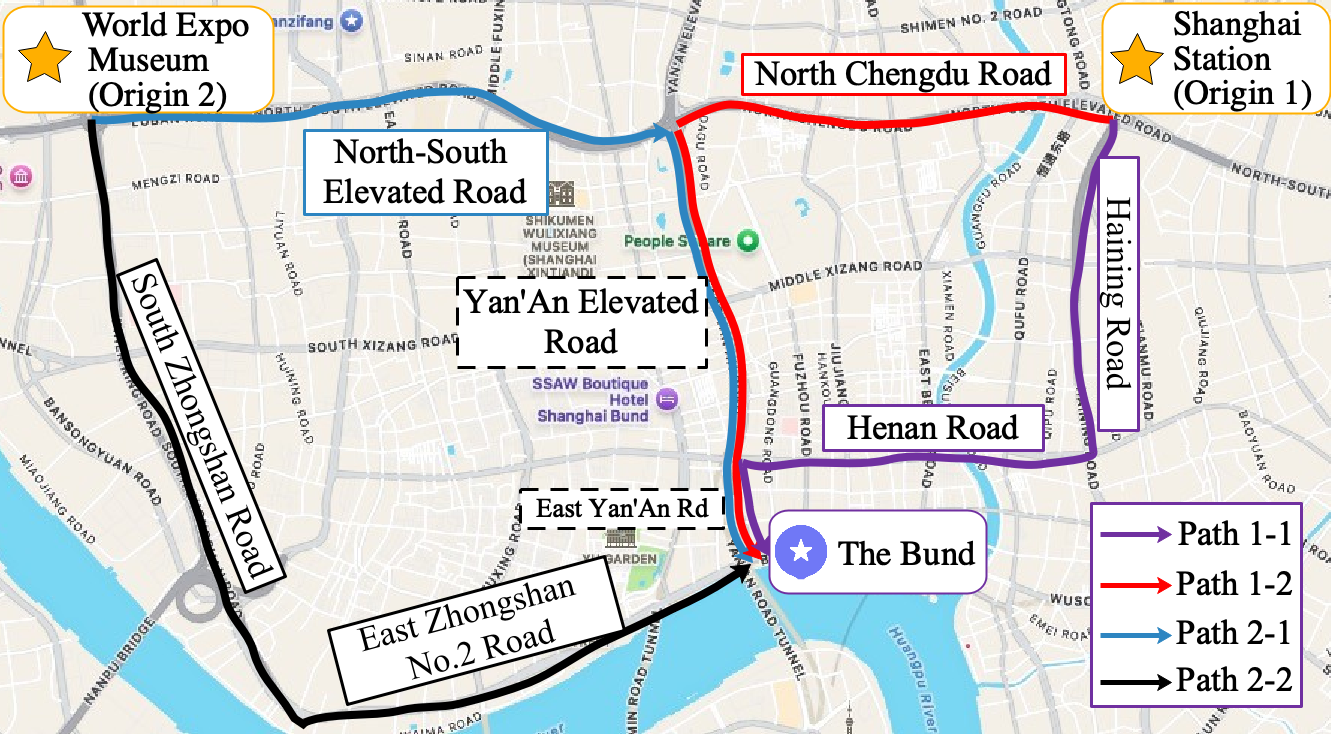}
    \caption{A hybrid road network with two origins, World Expo Museum and Shanghai Station, and a single destination, the Bund, Shanghai's most popular shopping and sightseeing area. Users randomly leave from World Expo Museum and Shanghai Station each day, and subsequently choose a path to reach the Bund.}
    \label{fig:sg_road}
\end{figure}

\section{Experiment Validation Using Real Datasets}\label{section6}

In this section, we conduct extensive experiments to evaluate our UPR mechanism's average performance versus the information sharing mechanism (used by Google Maps and Waze), the latest hiding mechanism in routing game literature (e.g., \cite{tavafoghi2017informational,farhadi2022dynamic,li2019recommending}), and the social optimum in terms of the average long-term social cost. 

Without much loss of generality, we focus on populous cities exhibiting time-varying road conditions and experiencing heavy traffic congestion. As such, we use live traffic datasets from Baidu Maps \cite{Baidumap} and real-time traffic speed data during similar peak hours to train our traffic model practically. Building on our parallel transportation network in Fig.~\ref{fig:congestion_game}, we extend it to a hybrid network by introducing two origin-destination pairs-Shanghai Station (origin 1) to the Bund and World Expo Museum (origin 2) to the Bund-with overlapping paths (i.e., Yan'An Elevated Road and East Yan'An Road) in Fig.~\ref{fig:sg_road}. This setup captures the typical flow of visitors traveling from origin 1 and origin 2 to the Bund.

For the route from origin 1 to the Bund, the options are:
\begin{itemize}
    \item Path 1-1: via First Haining Road, Henan Road, and then East Yan'An Road.
    \item Path 1-2: via North Chengdu Road, Yan'An Elevated road, and East Yan'An Road.
\end{itemize}
For the route from origin 2 to the Bund, the options are:
\begin{itemize}
    \item Path 2-1: via North-South Elevated Road, Yan'An Elevated Road, and East Yan'An Road.
    \item Path 2-2: via South Zhongshan Road and East Zhongshan No.2 Road.
\end{itemize}

The mean arrival rates are $\alpha^1=102$ and $\alpha^2=56$ cars every $5$ minutes at origins origin 1 and origin 2, respectively, with a standard deviation of $5.62$. To develop a practical congestion model with travel costs, we mine and analyze extensive data on the SpeedBand values across the eight roads, which fluctuate every $5$ minutes in the peak-hour periods. 
Our data analysis verifies that the traffic conditions on North Chengdu Road, Yan'An Elevated Road, and East Yan'An Road in path 1-2, as well as North-South Elevated Road in path 2-1, can be well approximated as a Markov chain with two discretized states (high and low traffic states) for learning, as depicted in Fig.~\ref{fig:mdp}. 
In contrast, Haining Road, and Henan Road in path 1-1, as well as South Zhongshan Road and East Zhongshan No.2 Road in path 2-2, display more deterministic conditions.
Following similar methodologies from \cite{eddy1998profile,chen2016predicting,saad2019whitespace}, we train the hidden Markov model (HMM) and obtain the transition probability matrices $\text{N\_CD}, \text{YA\_E}, \text{E\_YA}$, and $\text{NS\_E}$ for North Chengdu Road, Yan'An Elevated Road, East Yan'An Road, and North-South Elevated Road, respectively:
\begin{align*}
    \text{N\_CD}=\begin{bmatrix}
        0.6239 & 0.3761\\
        0.3525 & 0.6475
    \end{bmatrix}&,\text{YA\_E}=\begin{bmatrix}
        0.7873 & 0.2127\\
        0.3039 & 0.6961
    \end{bmatrix},\\
    \text{E\_YA}=\begin{bmatrix}
        0.8443 & 0.1557\\
        0.1326 & 0.8674
    \end{bmatrix}&,
    \text{NS\_E}=\begin{bmatrix}
        0.8730 & 0.1280\\
        0.0898 & 0.9002
    \end{bmatrix}.
\end{align*}
In this hybrid two-origin network, users assess the travel latency as the travel cost. Initially, the four stochastic roads' hazard beliefs satisfy $x_{\text{NCD}}(0)=0.5$, $x_{\text{YAE}}(0)=0.6$, $x_{\text{EYA}}(0)=0.7$ and $x_{\text{NSE}}(0)=0.3$ before the peak hours.

Under both sharing and hiding mechanisms, users at origin 1 estimate the combined travel costs of North Chengdu Road and Yan'An Elevated Road as $\mathbb{E}[c_{1-2}(t)]$ for path 1-2, incorporating the congestion cost from the expected flow of users on path 2-1. Since paths 1-1 and 1-2 share East Yan'An Road, users at origin 1 compare $\mathbb{E}[c_{1-2}(t)]$ with the fixed cost $c_{1-1}$ (covering the combined travel costs of Haining Road and Henan Road) to make routing decisions. 
Similarly, users at origin 2 estimate the combined travel costs of North-South Elevated Road, Yan'An Elevated Road, and East Yan'An Road for path 2-1 as $\mathbb{E}[c_{2-1}(t)]$. Then they compare $\mathbb{E}[c_{2-1}(t)]$ and the fixed cost $c_{2-2}$ to determine, based on self-interest, which path minimizes their personal travel costs.
While the socially optimal policy continuously search for the path that minimizes the long-term social cost for all users. To apply our UPR mechanism in \Cref{def:UPR}, we combine the three roads on paths 1-2 and 2-1 and update the parameters in (\ref{pr(pi=i)_multi}) to provide probabilistic recommendations. These recommendations direct users with differential information ages toward specific paths from origins origin 1 and origin 2, based on the public hazard beliefs of the four roads as in (\ref{pr(pi=i)_multi}).
In addition, we set the discount factor $\rho=0.95$ for the long-term social costs and $\epsilon=1$ as the minimum recommended flow in (\ref{fm(t)}), (\ref{f*(t)}), and (\ref{pr(pi=i)_multi}). 
After training a practical traffic model and applying mechanisms, we are ready to conduct experiments to show our UPR mechanism's average performance versus the other mechanisms.
We run $100$ experiments (each with $30$ time slots or equivalently $150$ minutes) for averaging the long-term social costs during peak hours.

\begin{figure}[t]
    \captionsetup{font={footnotesize}}
    \centering
    \includegraphics[width=0.4\textwidth]{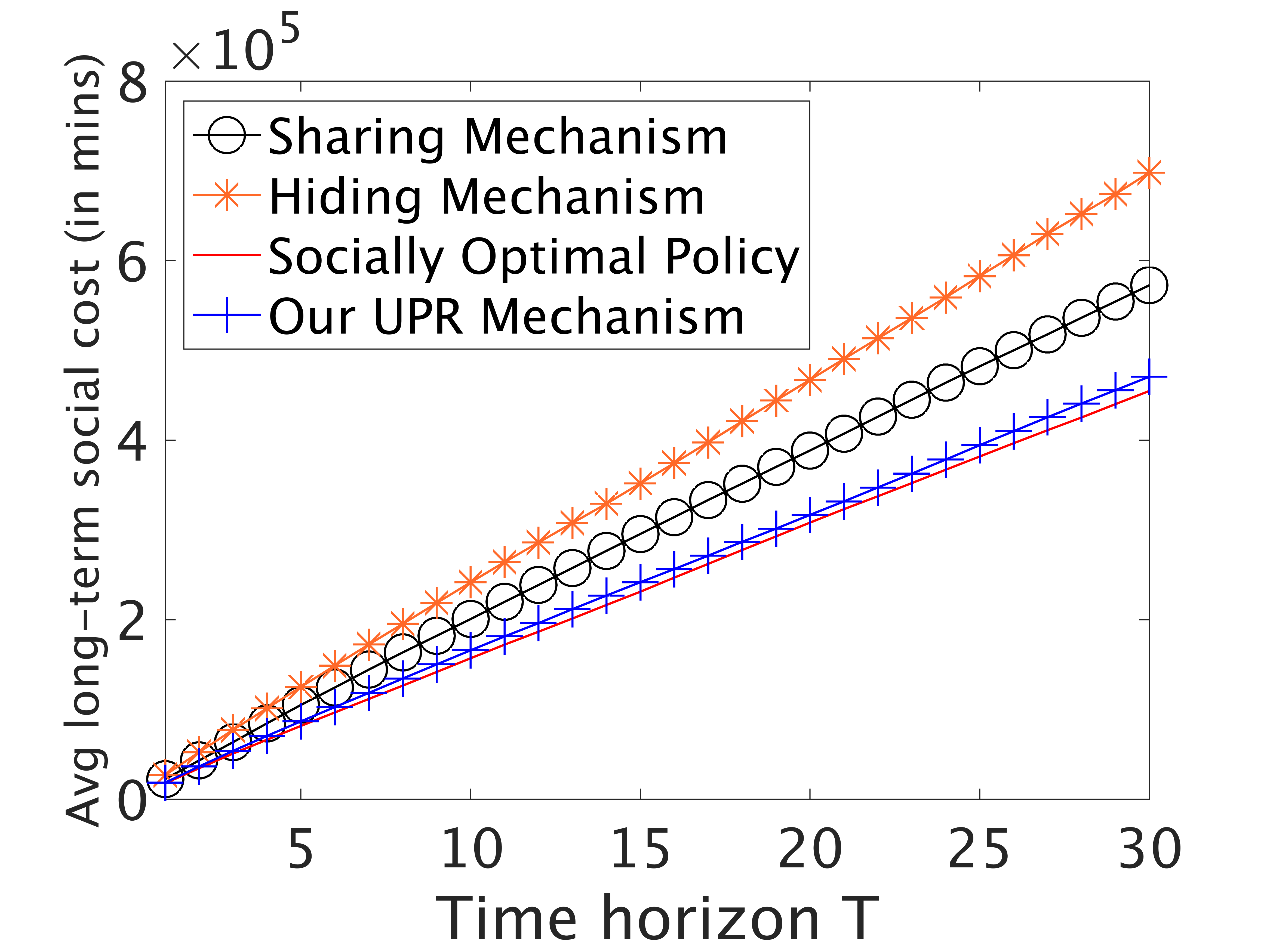}
    \caption{Average long-term social costs (in minutes) under information sharing, hiding, the optimum, and our UPR mechanism versus time horizon $T$.}
    \label{fig:avg_cost}
\end{figure}

Fig. \ref{fig:avg_cost} compares the long-term social cost performances of the information sharing, hiding, the optimum, and our UPR mechanisms versus time horizon $T$. It demonstrates that our UPR mechanism has less than $5\%$ efficiency loss from the social optimum for any time horizon $T$, while the sharing and hiding mechanisms have around $20\%$ and $40\%$ efficiency losses, respectively.

\section{Conclusions}\label{section7}
In this paper, we study how to regulate human-in-the-loop learning (HILL) in repeated routing games, where we face non-myopic users of differential past observations and need new mechanisms (preferably non-monetary) to persuade users to adhere to the optimal path recommendations. We first prove that no matter under the information sharing mechanism in use or the latest routing literature's hiding mechanism, the resultant price of anarchy (PoA) for measuring the efficiency loss from social optimum can approach infinity, telling arbitrarily poor exploration-exploitation tradeoff over time. Then we propose a novel user-differential probabilistic recommendation (UPR) mechanism to differentiate and randomize path recommendations for users with differential learning histories. 
We prove that our UPR mechanism ensures interim individual rationality for all users and significantly reduces $\text{PoA}=\infty$ to close-to-optimal $\text{PoA}=1+\frac{1}{4N+3}$, which cannot be further reduced by any other non-monetary mechanism. In addition to theoretical analysis, we conduct extensive experiments using real-world datasets to generalize our routing graphs and validate the close-to-optimal UPR performance mechanism.





\newpage

\balance

\begin{IEEEbiography}[{\includegraphics[width=1in,height=1.25in,clip,keepaspectratio]{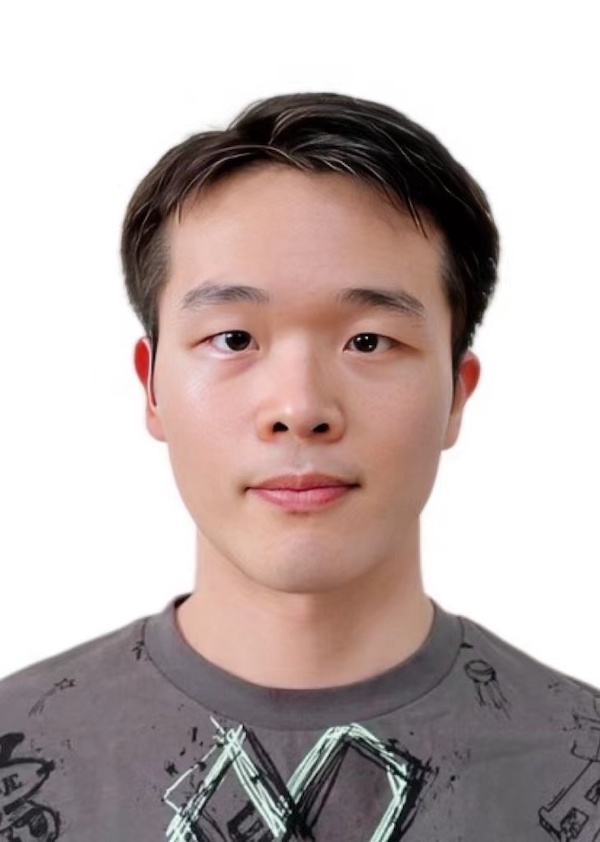}}]{Hongbo Li}(S'24-M'24)
received the Ph.D. degree from Singapore University of Technology and Design (SUTD) in 2024. He is currently a Postdoctoral Research Fellow with the Pillar of Engineering Systems and Design, SUTD. In 2024, he was a Visiting Scholar at The Ohio State University (OSU), Columbus, OH, USA. His research interests include networked AI, game theory and mechanism design, and machine learning theory.
\end{IEEEbiography}

\vspace{11pt}
\begin{IEEEbiography}[{\includegraphics[width=1in,height=1.25in,clip,keepaspectratio]{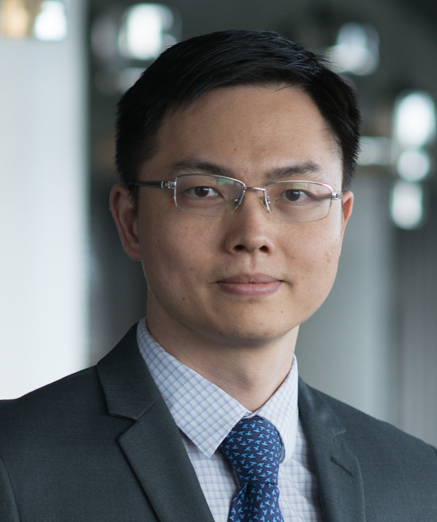}}]{Lingjie Duan}(S'09-M'12-SM'17) received the Ph.D. degree from The Chinese University of Hong Kong in 2012. He is an Associate Professor at the Singapore University of Technology and Design (SUTD) and is an Associate Head of Pillar (AHOP) of Engineering Systems and Design. In 2011, he was a Visiting Scholar at University of California at Berkeley, Berkeley, CA, USA. His research interests include network economics and game theory, network security and privacy, energy harvesting wireless communications, and mobile crowdsourcing. He is an Associate Editor of IEEE/ACM Transactions on Networking and IEEE Transactions on Mobile Computing. He was an Editor of IEEE Transactions on Wireless Communications and IEEE Communications Surveys and Tutorials. He also served as a Guest Editor of the IEEE Journal on Selected Areas in Communications Special Issue on Human-in-the-Loop Mobile Networks, as well as IEEE Wireless Communications Magazine. He is a General Chair of WiOpt 2023 Conference and is a regular TPC member of some other top conferences (e.g., INFOCOM, MobiHoc, SECON). He received the SUTD Excellence in Research Award in 2016 and the 10th IEEE ComSoc Asia-Pacific Outstanding Young Researcher Award in 2015. 
\end{IEEEbiography}

\appendix

\subsection{Proof of Lemma 1}
We consider a typical two-path network with one deterministic path (path 0) and one stochastic path (path 1) to derive the three cases of (7) in Lemma 1, depending on the relationship between cost $c_0$ of deterministic path 0 and $\mathbb{E}[c_1(t)|x_1(t)]$ of stochastic path 1. 

Under the information-sharing mechanism, a selfish user will always choose the path that minimizes his own long-term cost at any $t$. 
Define $Q^{(s)}\big(x_1(t)|f_1^{(s)}(x_1(t))\big)$ to be the minimal long-run expected cost of a single user, given $f_1^{(s)}(x_1(t))$ flow of users choosing path 1 except for himself. Denote this user's selfish decision under the sharing mechanism by $\pi^{(s)}(t)$. We obtain:
\begin{align}
    &Q^{(s)}\big(x_1(t)|f_1^{(s)}(x_1(t))\big)=\tag{23}\label{Q_s} \\ &\min\Big\{c_0+1-f_1^{(s)}(x_1(t))+\rho Q_0^{(s)}\big(x_1(t+1)|f_1^{(s)}(x_1(t))\big),\notag \\ 
    &\mathbb{E}[c_1(t)|x_1(t)]+f_1^{(s)}(x_1(t))+\rho Q_1^{(s)}\big(x_1(t+1)|f_1^{(s)}(x_1(t))\big)\Big\}\notag
\end{align}
by estimating $f_1^{(s)}(x_1(t))$ under $x_1(t)$, where $Q_0^{(s)}\big(x_1(t+1)|f_1^{(s)}(x_1(t))$ and $Q_1^{(s)}\big(x_1(t+1)|f_1^{(s)}(x_1(t))$ are cost-to-go functions under $\pi^{(s)}(t)=0$ and $\pi^{(s)}(t)=1$, respectively.

From (\ref{Q_s}), if $f_1^{(s)}(x_1(t))$ is obviously greater than $0$ under $\mathbb{E}[c_1(t)|x_1(t)]< c_0+1$, then $Q_0^{(s)}\big(x_1(t+1)|f_1^{(s)}(x_1(t))\big)=Q_1^{(s)}\big(x_1(t+1)|f_1^{(s)}(x_1(t))\big)$. This is because some other users will travel on stochastic path 1 and share the same observation there, and this user can always exploit the shared information on path 1 in the next time slot, no matter whether he chooses path $\pi^{(s)}(t)=1$ or not. In this case, he will only compare immediate costs of the two paths in (\ref{Q_s}) to make his decision. Therefore, at the Nash equilibrium, the immediate travel costs on the two paths satisfy
\begin{equation*}
    c_0+1-f_1^{(s)}(x_1(t))=\mathbb{E}[c_1(t)|x_1(t)]+f_1^{(s)}(x_1(t)).
\end{equation*}
Then the solution of $f_1^{(s)}(x_1(t))$ has the following three cases:
\begin{itemize}
    \item If $\mathbb{E}[c_1(t)|x_1(t)]\leq c_0-1$, $f_1^{(s)}(x_1(t))=1$ with the maximum flow on stochastic path 1, which is the second case of (7). 
    \item If $c_0+1>\mathbb{E}[c_1(t)|x_1(t)]> c_0-1$, we obtain $f_1^{(s)}(x_1(t))=\frac{1}{2}+\frac{c_0-\mathbb{E}[c_1(t)|x_1(t)]}{2}$, which is the third case of (7).
    \item On the other hand, if $\mathbb{E}[c_1(t)|x_1(t)]\geq c_0+1$, $f_1^{(s)}(x_1(t))$ may be zero without other users choosing path~1. In this case, $Q_0^{(s)}\big(x_1(t+1)|0\big)\neq Q_1^{(s)}\big(x_1(t+1)|\epsilon\big)$, as there will be no information update if this user also chooses deterministic path 0. Then if $Q_0^{(s)}\big(x_1(t+1)|0\big)> Q_1^{(s)}\big(x_1(t+1)|\epsilon\big)$, this user will choose stochastic path 1 for lower long-term cost, and thus $f_1^{(s)}(x_1(t))=\epsilon$. Otherwise, no user will choose stochastic path 1, and $f_1^{(s)}(x_1(t))=0$. In summary, $f_1^{(s)}(x_1(t))=\epsilon\cdot \mathbf{1}\{Q^{(s)}(x_1(t)|\epsilon)\leq Q^{(s)}(x_1(t)|0)\}$, which is the first case of (7). 
\end{itemize}
This finishes the proof of decision-making at the two-path network.

For an arbitrary number $N\geq 2$, if the travel cost of a stochastic path $i$ is larger than other paths, a user also compares his long-term costs $Q_0^{(s)}(\mathbf{x}(t)|f_i^{(s)}(\mathbf{x}(t))=0)$ and $Q_1^{(s)}(\mathbf{x}(t)|f_i^{(s)}(\mathbf{x}(t))=\epsilon)$. While in other cases, he will balance the immediate cost on path $i$ with other paths to reach the Nash equilibrium.

\subsection{Proof of Lemma 2}
We still consider the typical two-path network to derive $f_{1,1}^{\emptyset}(y_{1,1}(t))$ and $f_{1,2}^{\emptyset}(y_{1,2}(t))$ for users with information age $d_1(t)=1$ and $d_1(t)=2$, respectively. 

Given $f_1^{\emptyset}(\mathbf{y}(t-1))$ flow of users exploring path $i$ in the last time, the current flow of users with $d_1(t)=1$ is thus $f_1^{\emptyset}(\mathbf{y}(t-1))$. As they have the latest information on this path with $d_1(t)=1$, they will infer other users' private belief $y_{i,2}(t)$ with $d_1(t)=2$ according to (4) and the exact flow $f_{i,2}^{\emptyset}(y_{i,2}(t))$ of these users on path~$i$. Then they decide $f^{\emptyset}_{i,1}(y_{i,1}(t))$ to minimize their own costs. Thus, we first derive $f^{\emptyset}_{i,2}(y_{i,2}(t))$ for users with $d_1(t)=2$. 

For the $\big(1-f_1^{\emptyset}(\mathbf{y}(t-1))\big)$ flow of users with $d_1(t)=2$, they only hold rough belief $y_{i,2}(t)$ to decide $f^{\emptyset}_{i,2}(y_{i,2}(t))$ to minimize their expected long-run costs. To realize the expected flow $f_1^{(s)}(y_{1,2}(t))$ at the Nash equilibrium, the probability for each of them choosing stochastic path 1 is thus $f_1^{(s)}(y_{1,2}(t))$, which is derived by (7). Therefore, the routing flow of users with $d_1(t)=2$ on path 1 is:
\begin{align*}
    f^{\emptyset}_{1,2}(y_{1,2}(t))=(1-f_1^{\emptyset}(\mathbf{y}(t-1)))f_1^{(s)}(y_{1,2}(t)).
\end{align*}

We now go back to users with minimum information age $d_1(t)=1$ to decide $f^{\emptyset}_{1,1}(y_{1,1}(t))$. As these users know exact $x_1(t)$ of path 1, they aim to realize routing flow $f^{(s)}_1(y_{1,1}(t))$ on path 1 at the Nash Equilibrium. Given $f^{\emptyset}_{1,2}(y_{1,2}(t))$ flow of users choosing stochastic path~1 in (13) already, they will compare $f^{(s)}_1(y_{1,1}(t))$ with $f^{\emptyset}_{1,2}(y_{1,2}(t))$ to make decisions:
\begin{itemize}
    \item If $f^{\emptyset}_{1,2}(y_{1,2}(t))\geq f^{(s)}_1(y_{1,1}(t))$, none of users with $d_1(t)=1$ will choose path 1.
    \item Otherwise, there will be $f^{(s)}_1(y_{1,1}(t))-f^{\emptyset}_{1,2}(y_{1,2}(t))$ flow of users with $d_1(t)=1$ choosing path 1. This flow is upper bounded by the total flow of users with minimum age $d_1(t)=1$, i.e., $f_1^{\emptyset}(\mathbf{y}(t-1))$.
\end{itemize}
In summary, we finally obtain $f^{\emptyset}_{1,1}(y_{1,1}(t))$ in (10).
Given $f^{\emptyset}_{1,1}(y_{1,1}(t))$ in (12) and $f^{\emptyset}_{1,2}(y_{1,2}(t))$ in (13), the total flow of users finally choose path 1 is $f^{\emptyset}_{1,1}(y_{1,1}(t))+f^{\emptyset}_{1,2}(y_{1,2}(t))$. 

\subsection{Proof of Lemma 3}

In (15), even if the expected travel cost $\mathbb{E}[c_i(t)|x_i(t)]$ on path~$i$ is much higher than others, the socially optimal policy may still recommend a small flow $\epsilon> 0$ of users to learn $x_i(t)$ on path~$i$. As $\epsilon\rightarrow 0$, this infinitesimal flow of users' immediate total cost $\epsilon(\mathbb{E}[c_i(t)|x_i(t)]+\epsilon)$ is negligible. At the same time, their learned information on path $i$ can help reduce future costs for all users. Therefore, the socially optimal policy always decides $f_i^*(t)\geq \epsilon$ for any stochastic path~$i$. 

As there are always some users traveling on stochastic path $i$ to update $x_1(t+1)$, according to (15), the socially optimal policy only needs to minimize the immediate social cost $c^*(x_1(t))$, which is defined in (8). To derive the optimal flow $f_1^*(x_1(t))$, we calculate the first-order-derivative of $c^*(x_1(t))$:
\begin{align*}
    \frac{\partial c^*(x_1(t))}{f_1^*(x_1(t))}=4f_1^*(x_1(t))+(\mathbb{E}[c_1(t)|x_1(t)]-c_0-2).
\end{align*}
Then there are also three cases for deciding the optimal routing flow $f_1^*(x_1(t))$:
\begin{itemize}
    \item If $\mathbb{E}[c_1(t)|x_1(t)]\leq c_0-2$, $\frac{\partial c^*(x_1(t))}{f_1^*(x_1(t))}\leq 0$ is always true. In this case, $f_1^*(x_1(t))=1$, which is the second case of (16).
    \item If $\mathbb{E}[c_1(t)|x_1(t)]\geq c_0+2$, $\frac{\partial c^*(x_1(t))}{f_1^*(x_1(t))}\geq 0$ is always true. In this case, $f_1^*(x_1(t))=\epsilon$, which is the first case of (16). 
    \item Otherwise, by solving $\frac{\partial c^*(x_1(t))}{f_1^*(x_1(t))}=0$, we obtain $f_1^*(x_1(t))=\frac{1}{2}+\frac{c_0-\mathbb{E}[c_1(t)|x_1(t)]}{4}$, which is the third case of (16). 
\end{itemize} 
This completes the proof of Lemma 3.

\subsection{Proof of Lemma 4}
We first prove that $x_i(t+1)$ increases monotonically with $x'_i(t)$ in (2). Then we prove $x_i(t+1)$ is lower- and upper-bounded by $q_{LH}$ and $q_{HH}$ at $x'_i(t)=0$ and $x'_i(t)=1$, respectively. Finally, according to the definitions of $y_{i,1}(t)$ in (3) and $y_{i,2}(t)$ in (4), we can obtain the same bounds for $y_{i,d_i(t)}(t)$.

According to (2), we obtain the first-order-derivative of $x_i(t+1)$ w.r.t. $x'_i(t)$ below:
\begin{align*}
    \frac{\partial x_i(t+1)}{\partial x'_i(t)}=q_{HH}-q_{LH},
\end{align*}
which is always greater than $0$ based on the Markov chain in Figure 1b. If $x_i'(t)=0$ with a good observation, then $x_i(t+1)=q_{LH}$, which is the minimum. While if $x_i'(t)=1$, we obtain the maximum hazard belief $x_i(t+1)=q_{HH}$. 

Therefore, public belief $x_i(t)$ belongs to the range $[q_{LH}, q_{HH}]$ for any $t$. And private belief $y_{i,d_i(t)}(t)$ holds the same bounds.

\subsection{Proof of Proposition 1}
We will prove this lower bound of PoA in (18) by showing the worst-case scenario.

At the initial time $t=0$ with $\mathbb{E}[c_i(0)|x_i(0)]\rightarrow c_0+1$, according to (7), a flow of selfish users will choose to explore each stochastic path $i$. As there are $N$ stochastic paths, there is a probability $1-\prod_{i=1}^Nx_i(0)$ probability for them to find at least one stochastic path $i$ with $c_i(0)=c_L$. In this case, the expected social cost caused by selfish users satisfies
\begin{align}
    &C^{(s)}(\mathbf{x}(0))\notag\\=&(c_0+1-\epsilon)(1-\epsilon)+N\mathbb{E}[c_i(0)|x_i(0)]\epsilon+\rho C^{(s)}(\mathbf{x}(1))\notag\\
    \geq &c_0+1+\sum_{j=1}^{\infty}\rho^j\Big( c_0+1 \prod_{i=1}^N x_i(0)+\frac{1}{N}(1-\prod_{i=1}^N x_i(0))\Big)\notag\\
    \geq&c_0+1+\sum_{j=1}^{\infty}\rho^j\Big((c_0+1) \prod_{i=1}^N x_i(0)\Big) \notag\\
    =&c_0+1+\frac{\rho(c_0+1)\prod_{i=1}^N x_i(0)}{1-\rho}.\tag{24}\label{C^s_poa}
\end{align}
Here the first inequality is due to
\begin{align*}
    C^{(s)}(\mathbf{x}(1))\geq \sum_{j=0}^{\infty}\rho^j\Big((c_0+1) \prod_{i=1}^N x_i(0)+\frac{1}{N}(1-\prod_{i=1}^N x_i(0))\Big),
\end{align*}
which is because selfish users will never explore any path~$i$ with probability $\prod_{i=1}^N x_i(0)$ after finding $x_i(t)=q_{HH}$ there, and $C^{(s)}(\mathbf{x}(1))$ is lower bounded by the best case that all users keep exploiting all $N$ stochastic paths with probability $1-\prod_{i=1}^N x_i(0)$. Note that $\mathbb{E}[x_i(t)]=x_i(0)$ according to (2) for any time $t\geq 1$. 

Under selfish social cost $C^{(s)}(\mathbf{x}(1))$ in (\ref{C^s_poa}), we next calculate the optimal social cost. Different from the minimum exploration of the selfish policy, the socially optimal policy always lets an $\epsilon$ flow of users explore any stochastic path~$i$. Even though they find bad traffic conditions on path~$i$, there will still be an $\epsilon$ flow exploring this path until find good conditions. If users observe $c_i(t)=c_H$ on path $i$, let $k$ denote the maximum time slots before observing good condition $c_i(t+k)=c_L$ again on path $i$, which satisfies $q_{HH}^k\leq \delta$, where $\delta\rightarrow 0$ is a small infinitesimal. Here $x_i(t)$ becomes $q_{HH}$ after observing $c_H$ on path $i$, according to Lemma~4. Then we obtain $k=\lceil \log_{q_{HH}}\delta\rceil$. After observing $c_L$ again, all users can exploit this path for near-to-zero cost there. As there are $N$ paths, after at most $\frac{k}{N}$ time slots, users will observe at least one path with a good condition.
Given $q_{LL}\rightarrow 1$ and $\frac{1-q_{LL}}{1-q_{HH}}\rightarrow 0$, we can finally obtain the minimum social cost based on the above analysis:
\begin{align}
    &C^*(\mathbf{x}(0))\notag\\ \leq& c_0+1+(1-\prod_{i=1}^Nx_i(0))\sum_{j=1}^{\infty} \rho^j +\prod_{i=1}^N x_i(0)\rho C^*(q_{HH}) \notag\\ \leq&c_0+1+(1-\prod_{i=1}^Nx_i(0))\frac{\rho}{1-\rho} +\prod_{i=1}^Nx_i(0)\sum_{\tau=1}^{\frac{\lceil \log_{q_{HH}}\delta \rceil}{N}}\rho^{\tau}(c_0+1)\notag\\ =&c_0+1+(1-\prod_{i=1}^Nx_i(0))\frac{\rho}{1-\rho}+\tag{25}\label{C^*_poa} \\&\frac{\rho-\rho^{\frac{\lceil \log_{q_{HH}}\delta \rceil}{N}}(c_0+1)}{1-\rho}\prod_{i=1}^Nx_i(0).\notag
\end{align}

Combing the social cost $C^{(s)}(\mathbf{x}(0))$ caused by the selfish players in (\ref{C^s_poa}) above and the minimum social cost $C^*(\mathbf{x}(0))$ in (\ref{C^*_poa}), we obtain
\begin{align*}
    &\text{PoA}^{(s)}\\ \geq& \frac{C^{(s)}(\mathbf{x}(0))}{C^*(\mathbf{x}(0)}\\ 
    =& \Scale[1.2]{\frac{1-\rho+\rho \prod_{i=1}^Nx_i(0) }{1-\rho +(1-\prod_{i=1}^Nx_i(0))\rho\frac{1}{c_0+1}+(\rho-\rho^{\frac{\lceil \log_{q_{HH}}\delta \rceil}{N}})\prod_{i=1}^Nx_i(0)}}\\ =&\frac{1-\rho+\rho \prod_{i=1}^Nx_i(0)}{1-\rho +(\rho-\rho^{\frac{\lceil \log_{q_{HH}}\delta \rceil}{N}})\prod_{i=1}^Nx_i(0)},
\end{align*}
as $c_0\gg 1$ to make $\frac{1}{c_0+1}=0$.

This PoA lower bound approaches infinity as $\rho^{\frac{\lceil \log_{q_{HH}}\delta \rceil}{N}}\rightarrow 1$ under $\mathcal{O}(\frac{1}{1-\rho})\gg \mathcal{O}(\frac{1}{1-q_{HH}})$.

\subsection{Proof of Proposition 2}
To prove Proposition 2, we consider the same worst-case scenario with minimum exploration as the proof in Appendix~E.

Under the information-hiding mechanism, according to (12) and (13), there will be a small $\epsilon$ flow of users choosing each stochastic path $i$ to travel. 
For this $\epsilon$ flow of users, if they observe $c_i(0)=c_L$ on path $i$, they still choose this path at the next time slot $t+1$. If this $f_i^{\emptyset}(0)=\epsilon$ flow of users observe $c_i(0)=c_H$ there, none of them will go to this path again since $t=1$, based on $f_{i,1}^{\emptyset}(y_{i,1}(t))$ in (12). For the other $1-\epsilon$ flow of users choosing path 0 at time $t=0$, there is no observation to update their posterior belief $y_{i,2}(1)$ for the next time slot $t=1$. 
Thus, at time $t=1$, these users with information age $d_i(1)=2$ still choose path 0. 
However, at the end of time $t=1$, these users can observe $f^{\emptyset}_i(1)$ to update their beliefs $y_{i,2}(2)$:
\begin{itemize}
    \item If they observe $f^{\emptyset}_i(1)=0$, they infer that $c_i(0)=c_H$, and they keep choosing path 0 since the next time slot $t=2$.
    \item If they observe $f^{\emptyset}_i(1)=\epsilon$, they infer that $c_i(0)=c_L$, and they will choose stochastic path $i$ with good conditions since $t=2$.
\end{itemize}

Based on the analysis above and our proof of Proposition~1 in Appendix E, we calculate the social cost under the hiding mechanism as:
\begin{align*}
    C^{\emptyset}(\mathbf{x}(0))\geq& c_0+1+\rho (c_0+1)+\\ &\sum_{j=2}^{\infty}\rho^j\Big( (c_0+1) \prod_{i=1}^N x_i(0)+\frac{1}{N}(1-\prod_{i=1}^N x_i(0))\Big)\\
    \geq& (1+\rho)(c_0+1)+\frac{\rho^2(c_0+1)\prod_{i=1}^N x_i(0)}{1-\rho},
\end{align*}
where $f_i^{\emptyset}(1)=\epsilon$ such that the immediate cost for all users is $\rho(c_0+1)$ at time $t=1$.

For the socially optimal policy, it still always lets some users explore each stochastic path, and the caused minimum social cost $C^*(\mathbf{x}(0))$ is the same as (\ref{C^*_poa}) in Appendix E. Therefore, we calculate PoA under the hiding mechanism as
\begin{align*}
    \text{PoA}^{\emptyset}\geq \frac{1-\rho^2+\rho^2\prod_{i=1}^N x_i(0)}{1-\rho +(\rho-\rho^{\frac{k}{N}})\prod_{i=1}^Nx_i(0)},
\end{align*}
which approaches infinity under the same conditions as in Proposition 1.

\subsection{Proof of Lemma~5}

We demonstrate that the deterministic recommendation mechanism results in an infinite PoA by showing that users continue to make routing decisions as in the information-hiding mechanism.

Under the deterministic recommendation mechanism, when a user is advised to select a stochastic path, say $\pi(t)=i$, the only information it receives is $f_i^*(t)>0$. This limited information does not alter the user’s posterior belief $y_{i,d_i(t)}$ about path $i$, leading it to estimate its expected travel cost as it would without any recommendation. Consequently, the user disregards this recommendation. Similarly, if a user is advised to take safe path $\pi(t)=0$, its posterior beliefs regarding all other stochastic paths remain unaffected. Thus, each user continues to follow $f_i^{\emptyset}(\mathbf{y}(t))$ when deciding on a path, resulting in $\text{PoA}=\infty$, as established in Proposition~2.

\subsection{Proof of Proposition 3}
To prove Proposition 3, we consider the worst-case with maximum exploration on stochastic paths. 

Since users with minimum information age $d_i(t)=1$ have complete information of path $i$, any informational mechanism cannot change their path decisions. For example, we consider $q_{LH}\rightarrow 0, q_{HH}<1, c_L=0,x_i(t)=q_{LH}$ and $\mathbb{E}[c_i(t)|x_i(t)=q_{LH}]<c_0-1$. Users with $d_i(t)=1$ knows the exact hazard belief $x_i(t)=q_{LH}$ on path $i$. As the travel cost on path~$i$ satisfies $\mathbb{E}[c_i(t)|x_i(t)=q_{LH}]<c_0-1$, these users will always choose this path $i$ to exploit the low cost there. Motivated by this example, we will try to find the worst-case with the maximum exploration that every user has $d_i(t)=1$ to prove Proposition 3.

Let ${x_i(0)=q_{LH}}$, ${q_{LH}\rightarrow 0}$, ${q_{HH}<1}$, ${c_L=0}$ and ${\mathbb{E}[c_i(0)|x_i(0)]= c_0-\frac{1}{N}}$ for any stochastic path ${i\in \mathbf{N}}$. According to our analysis of ${f^{(s)}_i(\mathbf{x}(0))}$ in (7), there will be $\frac{1}{N}$ flow of users traveling on each stochastic path~$i$ at initial $t=0$. Then each user is aware of the public belief ${x_i(1)=q_{LH}}$ on his chosen stochastic path~$i$ with minimum information age ${d_i(1)=1}$ for next time ${t=1}$. Consequently, under ${\mathbb{E}[c_i(1)|x_i(1)]= c_0-\frac{1}{N}}$, they will behave the same as the sharing mechanism by continuing to choose their formerly chosen paths since $t= 1$, making any informational mechanism fail to regulate. The caused social cost by selfish users is 
\begin{align*}
    C^{(s)}(\mathbf{x}(0))&=\sum_{j=0}^{\infty} \rho^j(\mathbb{E}[c_i(1)|x_i(1)]+\frac{1}{N})\\&=\frac{\mathbb{E}[c_i(1)|x_i(1)]+\frac{1}{N}}{1-\rho}.
\end{align*}

On the other hand, according to (16) in Lemma 3, the socially optimal policy ensures the flow of users to choose deterministic path 0 as:
\begin{align*}
    {f^*(\mathbf{x}(t))}=\frac{1}{2(N+1)},
\end{align*}
to reduce the congestion cost on stochastic paths. Then we obtain the minimum social cost as:
\begin{align*}
    C^*(\mathbf{x}(0))=& \sum_{j=0}^\infty \rho^j \bigg(\frac{N}{2(N+1)} \Big(\mathbb{E}[c_i(0)|x_i(0)]+\frac{1}{2(N+1)}\Big)\\&+\Big(1-\frac{N}{2(N+1)}\Big)\Big(c_0+1-\frac{N}{2(N+1)}\Big)\bigg)\\
    =&\frac{4(N+1)(\mathbb{E}[c_i(1)|x_i(1)]+\frac{1}{N})}{(1-\rho)(4(N+1)-1)},
\end{align*}
which is derived by inputting $\mathbb{E}[c_i(1)|x_i(1)]+\frac{1}{N}=c_0$ into the right-hand side of the inequality.
Finally, the resulting minimal PoA is 
\begin{align*}
    \text{PoA}\geq \frac{C^{\emptyset}(\mathbf{x}(0))}{C^*(\mathbf{x}(0))}= \frac{4(N+1)}{4N+3}.
\end{align*}

Therefore, any informational mechanism cannot reduce PoA to less than $1+\frac{1}{4N+3}$.

\subsection{Proof of Theorem~1}
We first analyze the received recommendations of users with different private information under our UPR mechanism. Then, we prove that our UPR mechanism ensures IIR for all non-myopic users, prompting them to follow the system's random path recommendations. Finally, we demonstrate that under our UPR mechanism's random recommendations, users will only over-explore stochastic paths, thereby reducing the PoA to the minimum ratio (22), as stated in Proposition~3.

At any given time $t\geq 1$, our UPR provides differential recommendations to users, who have different past observations, to influence their path decisions:
\begin{itemize}
    \item [1.] For the $f_i^{(\text{UPR})}(t-1)$ flow of users who observed $c_i(t-1) = c_L$ on path $i$ with minimum age $d_i(t)=1$, any informational mechanism cannot stop their exploration on this path, based on our analysis in Proposition~3. 
    Therefore, the platform simply recommends that they follow their own selfish routing decisions in (7), and these users will stick to path $i$ at time $t$.
    \item [2.] For users with $d_i(t)=2$ on stochastic path $i$, the platform will provide probabilistic recommendations to let them choose any stochastic path $i\in I(t):=\{i\in \mathbb{N}|\Bar{f}^{(s)}(n,t)>f^{(\text{UPR})}_i(t-1)\}$, where $\Bar{f}^{(s)}(n,t)$ is the maximum flow of users on any path~$i$. 
    The condition in set $I(t)$ means that only if the flow of path $i$ in the last time slot satisfies $f^{(\text{UPR})}_i(t-1)<\Bar{f}^{(s)}(n,t)$, the platform will recommend some users with $d_i(t)=2$ to choose this path~$i$ in the current time slot $t$. If $f^{(\text{UPR})}_i(t-1)\geq \Bar{f}^{(s)}(n,t)$, our UPR mechanism will simply let the former $f^{(\text{UPR})}_i(t-1)$ flow of users with $d_i(t)=1$ follow their selfish decisions to determine $f^{(\text{UPR})}_i(t)=\Bar{f}^{(s)}(n,t)$, as shown in the case 1 above. 
\end{itemize}
In the second case, given $\Scale[0.95]{f^{(\text{UPR})}_i(t-1)<\Bar{f}^{(s)}(n,t)}$, if path $i$ is in a good condition, our UPR aims to realize $\Scale[0.95]{f^{(\text{UPR})}_i(t)=\Bar{f}^{(s)}(n,t)}$ flow of users exploiting this path. Conversely, if path $i$ is in a bad traffic condition, our UPR aims to ensure at least an $\epsilon$ flow of users choose this path, in order to achieve the social optimum in (16).
Note that if $f_i^{(s)}(\mathbf{x}(t))>\epsilon$ under $x_i(t)=q_{HH}$, in addition the recommended flow $\epsilon$, users with $d_i(t)=1$ will make their selfish decisions $f_i^{\emptyset}(\mathbf{x}(t))$ according to (12) to choose this path.

Based on the above analysis, we next prove that no matter receiving any private recommendation $\pi(t)$ under our UPR mechanism, each user is IIR to follow such recommendation. 

Suppose a user with $d_i(t)=2$ receives private path recommendation $\pi(t)=i$ for our UPR mechanism. According to (20), this user will infer the probability of a good condition on stochastic path $i$ as:
\begin{align}
    &\mathbf{Pr}(x_i(t)=q_{LH}|\pi(t)=i)\notag\\=&\frac{\mathbf{Pr}(x_i(t)=q_{LH},\pi(t)=i)}{\mathbf{Pr}(\pi_i(t)=i)}\notag\\=&\frac{\mathbf{Pr}(\pi_i(t)=i|x_i(t)=q_{LH})\mathbf{Pr}(x_i(t)=q_{LH})}{\sum_{x_i(t)\in\{q_{LH},q_{HH}\}}\mathbf{Pr}(\pi_i(t)=i|x_i(t))\mathbf{Pr}(x_i(t))}\notag\\=&\frac{ \frac{{\Bar{f}^{(s)}(n,t)-f^{(\text{UPR})}_i(t-1})}{1-f^{(\text{UPR})}_i(t-1)}\cdot (1-x_i(t-1))}{\frac{{\Bar{f}^{(s)}(n,t)-f^{(\text{UPR})}_i(t-1)}}{1-f^{(\text{UPR})}_i(t-1)}\cdot (1-x_i(t-1))+\epsilon\cdot x_i(t-1)},\label{tag_23}\tag{26}
\end{align}
which approaches $1$ as $\epsilon\rightarrow 0$. Here the first equality is derived by the Bayesian inference and the second equality is due to the law of total probability. Based on (\ref{tag_23}), for this player, it is almost certain that $x_i(t)=q_{LH}$. 

Then this user infers the expected travel cost on path $i$ as:
\begin{align*}
    &\lim_{\epsilon\rightarrow 0}\mathbb{E}[c_i(t)|\pi(t)=i]\\=&\lim_{\epsilon\rightarrow 0}\Big\{\mathbf{Pr}(x_i(t)=q_{LH}|\pi(t)=i)\mathbb{E}[c_i(t)|x_i(t)=q_{LH}]\\&+\mathbf{Pr}(x_i(t)=q_{HH}|\pi(t)=i)\mathbb{E}[c_i(t)|x_i(t)=q_{HH}]\Big\}\\
    =&\mathbb{E}[c_i(t)|x_i(t)=q_{LH}],
\end{align*}
which is the minimum expected cost. Based on our analysis that $f^{(\text{UPR})}_i(t-1)\leq \Bar{f}^{(s)}(n,t)$, following the recommendation $\pi(t)=i$ to choose stochastic path $i$ is IIR for this player. 

Similarly, if the user receives recommendation $\pi(t)=0$, it infers that any stochastic path $i$ is almost certainly with bad condition $x_i(t)=q_{HH}$. Suppose that other users follow our UPR recommendations. Then, for any stochastic path with good condition $x_i(t)=q_{LH}$, the expected flow of users on it will be $\Bar{f}^{(s)}(n,t)$. Therefore, following the recommendation to choose path 0 is individually rational.

As users always follow our UPR mechanism's recommendations, finally, we prove that our UPR realizes the minimum PoA in (22), which is the lower bound in Proposition 3 

According to recommendation probability $\mathbf{Pr}(\pi(t)=i|x_i(t))$ in (21), if stochastic path $i$ has a bad traffic condition $x_i(t)=q_{HH}$, our UPR recommends $(1-f_i^{(\text{UPR})}(t-1))\epsilon$ flow of users with information age $d_i(t)=2$ on path $i$ to choose this path. For the $f_i^{(\text{UPR})}(t-1)$ flow of users with $d_i(t)=1$, who follow (12) to make their selfish routing decisions, they reverse-engineer the recommended flow $\epsilon$ to make their routing decision $f_i^{\emptyset}(\mathbf{x}(t))$. In this case, the total flow of users choosing path $i$ under our UPR mechanism satisfies 
\begin{align}
    f^{(\text{UPR})}_i(t)&=(1-f_i^{(\text{UPR})}(t-1))\epsilon+f_i^{\emptyset}(\mathbf{x}(t))\notag\\
    &=\max\{f_1^{(s)}(y_{1,1}(t)),(1-f_i^{(\text{UPR})}(t-1))\epsilon\},\label{tag_24}\tag{27}
\end{align}
where the second equality is derived by (12).

Conversely, if path $i$ has a good traffic condition $x_i(t)=q_{LH}$, our UPR mechanism recommends an expected flow $\Bar{f}^{(s)}(n,t)-f^{(\text{UPR})}_i(t-1)$ of users with information age $d_i(t)=2$ on path $i$ to choose this path. As the other $f^{(\text{UPR})}_i(t-1)$ flow of users with $d_i(t)=1$ will stick to path $i$, our UPR mechanism achieves 
\begin{align}
    f^{(\text{UPR})}_i(t)=\Bar{f}^{(s)}(n,t).\label{tag_25}\tag{28}
\end{align}

Based on our analysis of (\ref{tag_24}) and (\ref{tag_25}), $f^{(\text{UPR})}_i(t)$ under our UPR mechanism always satisfies
\begin{align*}
    f^{(\text{UPR})}_i(t)\geq f_i^*(\mathbf{x}(t)),
\end{align*}
where $f_i^*(\mathbf{x}(t))$ is derived by (16) in Lemma~3. It means that our UPR mechanism always avoids users' under-exploration of any stochastic path $i$. Therefore, the infinite PoA's in Propositions 1 and 2 are successfully avoided. 

As users will only over-explore stochastic paths, the worst-case scenario is the maximum over-exploration as in Proposition~3, where all users choose stochastic paths such that $f^{(s)}(\mathbf{x}(t))=\Bar{f}^{(s)}(n,t)$. Since $x_i(t)=q_{LH}$ is always true, our UPR mechanism always recommends $f^{(\text{UPR})}_i(t)=\Bar{f}^{(s)}(n,t)$ flow of users on path $i$, based on (\ref{tag_25}). Therefore, our UPR mechanism leads to the same PoA $1+\frac{1}{4N+3}$ as in Proposition~3, which cannot be further reduced.

\vfill

\end{document}